\documentclass[pdftex,twocolumn,10pt,letterpaper]{extarticle}

\pdfoutput=1 

\newcommand{\AUTHORS}{Osama Haq\textsuperscript{1}, Cody Doucette\textsuperscript{2}, John W. Byers\textsuperscript{2}, 
and Fahad R. Dogar\textsuperscript{1}}

\newcommand{\TITLE}{Judicious QoS using Cloud Overlays}

\newcommand{\KEYWORDS}{Put your keywords here}
\newcommand{\CONFERENCE}{Somewhere}
\newcommand{\PAGENUMBERS}{yes}       
\newcommand{\COLOR}{yes}
\newcommand{\showComments}{yes}
\newcommand{\comment}[1]{}
\newcommand{\onlyAbstract}{no}

\newfont{\ttlfnt}{phvb8t at 18pt} 
\newfont{\aufnt}{phvr8t at 12pt}
\newfont{\auit}{phvro8t at 12pt}     
\newfont{\affaddr}{phvr8t at 10pt}

\usepackage[T1]{fontenc}
\usepackage[utf8]{inputenc}
\usepackage{verbatim}
\usepackage{newtxtext,newtxmath}       
\usepackage{bm}                        
\usepackage[scaled=0.92]{helvet}
\usepackage{courier}
\usepackage{subfigure}
\usepackage[ruled,noline,noend]{algorithm2e}
\usepackage{supertabular}

\usepackage[dvipsnames]{xcolor} 
\usepackage{wrapfig}

\usepackage{ifthen}
\ifthenelse{\equal{\PAGENUMBERS}{yes}}{%
\usepackage[nohead,
            left=1.0in,right=1.0in,top=1.0in,
            footskip=0.5in,bottom=1in,     
            columnsep=0.33in
		]{geometry}
}{%
\usepackage[noheadfoot,left=0.75in,right=0.85in,top=0.75in,
            footskip=0.5in,bottom=1in,
            columnsep=0.25in
	    ]{geometry}
}
\usepackage[font=bf]{caption}

\usepackage{titlesec}
\titlespacing{\paragraph}{0pt}{*1}{*1}      
\titleformat{\paragraph}[runin]{\normalfont\normalsize\bfseries}{\theparagraph}{1em}{}

\titleformat{\section}
  {\bf\large\uppercase}{\thesection.\quad}{0pt}{}
\titleformat{\subsubsection}
  {\it\large}{\thesubsubsection\quad}{0pt}{}

\usepackage{enumitem}
\setlist{itemsep=0pt,parsep=0pt}             

\usepackage{fancyhdr}

\ifthenelse{\equal{\PAGENUMBERS}{yes}}{%
  \pagestyle{plain}
}{%
  \pagestyle{empty}
}

\usepackage[numbers]{natbib}

\usepackage{booktabs}
\usepackage{multirow}
\usepackage{color}
\usepackage{colortbl}
\usepackage{float}                           

\ifthenelse{\equal{\COLOR}{yes}}{%
  \usepackage[colorlinks]{hyperref}
}{%
  \usepackage[pdfborder={0 0 0}]{hyperref}
}
\usepackage{url}

\hypersetup{%
pdfauthor = {\AUTHORS},
pdftitle = {\TITLE},
pdfsubject = {\CONFERENCE},
pdfkeywords = {\KEYWORDS},
bookmarksopen = {true}
}

\usepackage{hyperref}
\hypersetup{
     colorlinks   = true,
     citecolor    = blue,
     urlcolor = blue,
}

\definecolor{placeholderbg}{rgb}{0.85,0.85,0.85}

\usepackage[pdftex]{graphicx}
\usepackage{soul}
\soulregister\cite7
\soulregister\ref7
\soulregister\pageref7

\usepackage[absolute]{textpos}

\newfloat{acmcr}{b}{acmcr}

\newcommand{\note}[2]{
    \ifthenelse{\equal{\showComments}{yes}}{\textcolor{#1}{#2}}{}
}

\newcommand{\sys}{J-QoS} 
\newcommand{\rewan}{CR-WAN} 

\date{}
\title{\ttlfnt \TITLE}
\author{{\aufnt \AUTHORS \vspace{1ex}}\\
{\affaddr \textsuperscript{1}Tufts University, \textsuperscript{2}Boston University}}

\begin{document}

\maketitle


\begin{abstract}

We revisit the long-standing problem of providing network QoS to applications, and propose the concept of \emph{judicious} QoS --
combining the cheaper, best effort IP service with the cloud, which offers a highly reliable infrastructure and the ability to add in-network services, albeit at higher cost. Our proposed J-QoS framework offers a range of reliability services with different cost vs. delay trade-offs, including: i) a forwarding service that forwards packets over the cloud overlay, ii) a caching service, which stores packets inside the cloud and allows them to be pulled in case of packet loss or disruption on the Internet, and iii) a novel coding service that provides the least expensive packet recovery option by combining packets of multiple application streams and sending a small number of coded packets across the more expensive cloud paths. We demonstrate the feasibility of these services using measurements from RIPE Atlas and a live deployment on PlanetLab. We also consider case studies on how J-QoS works with services up and down the network stack, including Skype video conferencing, TCP-based web transfers, and cellular access networks.

\end{abstract}

\ifthenelse{\equal{\onlyAbstract}{no}}{%
\section{Introduction}
\label{sec:intro}

The limitations of IP's best effort service are well-known: it provides no guarantees on latency, packet loss, or bandwidth, which is restrictive, especially for interactive applications such as voice and video conferencing that require quality of service (QoS) support from the network.  Despite decades of research in this area, from ``first generation'' QoS proposals that required in-network changes (e.g., IntServ~\cite{intserv}) to overlay based solutions in the late 90's and early 2000's (e.g., RON~\cite{ron}, OverQoS~\cite{overqos}), an ideal solution still remains elusive -- a solution that can offer the reliability and performance of in-network solutions while being as easy to deploy as overlay-based solutions.  

Fortunately, the emergence of the cloud offers us an opportunity to revisit this problem. By the cloud, we refer to a distributed network of data centers (DCs), inter-connected through a private network (e.g., Azure, EC2, Google Cloud).  For any communication between two end-points, we can potentially use the cloud as an \emph{overlay}, with DCs acting as an insertion point for in-network services~\cite{rewanhotnets, via2016}. 
A cloud-based overlay offers unique opportunities: cloud paths are well-provisioned, offering low jitter and very high reliability; and each DC has visibility into many users and applications, so it can act as a unique vantage point for control and insertion of in-network services. On the flip side, using the cloud as an overlay can be costly: cloud providers charge for the use of their resources (e.g., processing, network connectivity), with wide area network (WAN) bandwidth being particularly expensive~\cite{swan, eurosys15, rewanhotnets}. Therefore, we argue that the most effective use of the cloud as an overlay is one that does so in a \emph{judicious} manner, in conjunction with the cheaper, best effort Internet paths.  

Toward this end, we present the Judicious QoS (\sys{}) framework, which uses the cloud infrastructure to provide enhanced network QoS. The main goal of \sys{} is to provide reliable and timely packet delivery to demanding applications.
\sys{} achieves this goal by offering three services with different cost vs. performance trade-offs.

The \emph{forwarding} service is the simplest one: it forwards packets over the cloud overlay, similar to how IP forwards packets on the Internet, but with the additional reliability and lower latency of cloud paths. A potential use case of this service is switching flows with consistently poor Internet paths onto the cloud overlay, similar to VIA~\cite{via2016}. The forwarding service acts as a building block for two new services, which use the storage and processing capability of the cloud, in addition to leveraging the high quality cloud paths.  

The \emph{caching} service provides (short term) storage of packets at a DC, leveraging the storage capability of the cloud, a functionality missing in IP routers. Unlike traditional caching approaches (e.g., CDNs), we use the cache for fast packet recovery. In our primary use case, a \emph{copy} of the packet is forwarded along the cloud path and cached at a DC close to the receiver. In case of a packet loss on the direct Internet path, the receiver retrieves the lost packet from its nearby DC's cache, rather than going all the way to the source. Because typical Internet path loss rates are low ($<1\%$), this reactive approach used by the caching service saves on the egress bandwidth of the cloud compared to the more pro-active forwarding service.

Finally, we present a novel \emph{coding} service, which provides the most economical option for protecting against packet losses on the best effort Internet paths, albeit at a slight increase in delay.
The coding service builds on top of the caching service but also leverages the processing capability of the cloud, in addition to its storage. It relies on the observation not all receivers experience a loss at the same time, so 
instead of caching the original packets at a DC near the receiver, we store a small number of coded packets. To recover a lost packet on the Internet, these coded packets are combined with data packets from other receivers, using an on-demand cooperative recovery process. 
For wide-area paths, this cooperative recovery process is still faster compared to relying on the source to retransmit a packet. The coding service, thus, exploits a number of observations and trends: encoding packets across users is made feasible because of cloud's visibility 
into many concurrent streams, and independent losses on Internet paths.
Similarly, cooperative recovery, using other receivers, can be feasible because of the low (and decreasing) latency between end-points and their nearby DCs.

While these services run in the cloud, \sys{} also provides suitable end-point support, which is important for fully leveraging the benefits of these services, including an API to access the services and a receiver driven loss detection mechanism. The API allows end-point applications to specify their latency budget, allowing \sys{} to choose the lowest cost service that would meet this requirement. The loss detection mechanism runs on the receiver; it predicts losses based on past packet arrivals and proactively undertakes loss recovery with the help of a suitable service running on a nearby DC (e.g., caching or coding).

We have implemented a prototype of \sys{} that logically sits just below the transport (e.g., TCP/UDP), providing enhanced reliability services on top of IP's best effort service. 
It seamlessly works with both TCP and UDP based applications (including encrypted traffic) without requiring any application modification, enabling a holistic evaluation of \sys{} along two broad themes: i) the feasibility and benefits of various \sys{} services in providing timely packet delivery, through measurements on RIPE Atlas~\cite{ripe-atlas}, and a deployment on the public cloud and PlanetLab~\cite{planetlab}, and ii) interplay of these services with protocols up and down the stack, with the help of case studies.

Our measurements on RIPE Atlas show the feasibility of our proposed services through latency measurements of the public Internet and the cloud overlay paths. For example, we observe that 80\% of the nodes can reach their nearest data center within 20ms, resulting in caching service recovering packets within a quarter of a round-trip time.

Our deployment on a public cloud for over a month helps us quantify the wide area performance improvement for PlanetLab paths. For example, our results show that the coding service is able to recover more than 70\% of losses, the recovery is typically within half a round-trip time, 
and the associated overhead of using the cloud judiciously is far less compared to other services (e.g., forwarding).  

Through case studies, we also evaluate how \sys{} interacts with protocols up and down the network stack -- we show that: i) \sys{}'s enhanced packet reliability can improve the user's QoE experience for a Skype video conferencing scenario, ii) \sys{} can speed up short web transfers by avoiding TCP timeouts and congestion avoidance caused by bursty losses, iii) under what scenarios it maybe feasible to use \sys{} on mobile networks, in terms of bandwidth, energy consumption, and latencies to nearby DCs.

\section{The Cloud as an Overlay}
\label{sec:motiv}
We consider using a cloud overlay as a potential solution to the network QoS problem. Some interactive applications, such as Skype and Google Hangouts, are already migrating their services to at least partial use of cloud relays~\cite{via2016}, but there has been little work in studying how to best utilize the cloud for such QoS sensitive applications. Therefore, we characterize the properties of cloud paths in terms of network conditions and cost, and ask:
can the cloud be leveraged in a cost-efficient way to make up for the Internet's performance limitations?

\paragraph{Benefits.} There are two key advantages of using the cloud as an overlay: 

\begin{enumerate} [leftmargin=*]
\item \textbf{Improved Performance.} 
Measurements show that cloud paths are highly reliable with a typical downtime target of a few minutes per month~\cite{goog-avail, cloudstudy-www}. A recent study shows that inter-DC paths have an order of magnitude lower loss rate, and significantly higher bandwidth, compared to public Internet paths~\cite{cloudstudy-www}. Similar benefits are being extended all the way up to ISP networks, with major cloud operators providing bandwidth-guaranteed pipes between their data centers and customer premises (e.g., Azure ExpressRoute~\cite{azure-expressroute}, AWS Direct Connect~\cite{aws-direct-connect}). These advances in the WAN as well as at the last hop are poised to make the \emph{entire} cloud overlay highly reliable with predictable latency between end-points.   

\item \textbf{Ability to insert in-network services.} The cloud infrastructure provides the ability to implement in-network services in software in a  scalable and fault tolerant fashion, with the help of network function virtualization (NFV)~\cite{combsekar2012,middleboxsherry}. 
These in-network services can leverage the storage and processing capabilities of the cloud to help with timely packet delivery, such as caching or coding of packets -- functionality that is infeasible to support in today's IP routers. 

\end{enumerate}

\paragraph{Cost.} Although the cloud as an overlay provides significant benefits, it can be expensive to use, especially due to the high cost of inter-DC bandwidth. 
Anecdotal evidence, as well as our discussions with operators, suggests that an inter-continental leased line could be an order of magnitude or more expensive compared to a connection to the best effort Internet.  
This reasoning underlies several recent proposals that try to make efficient use of inter-DC bandwidth in order to reduce their network costs~\cite{rewanhotnets,interdcpricing2016,netstitcher2011,movingbits-ifip2018}. 

\paragraph{Judiciously Using the Cloud.}
We argue that in current settings, we only need to rely on the cloud whenever the best-effort Internet cannot provide the desired QoS. 
For example, an application using an Internet path with 1\% loss is still getting 99\% of its packets delivered, so it could potentially make minimal use of the cloud and yet get its desired QoS.
To this end, we propose judicious use of cloud resources: 
leveraging the availability, performance, and other benefits of cloud only when the best-effort Internet fails to meet the desired QoS.
The key to our idea is to not only leverage the performance benefits of cloud paths (which other overlay proposals like VIA~\cite{via2016} propose), but to also judiciously use the storage and processing capability of the cloud, resulting in novel cloud-based services for timely packet delivery.

\begin{figure}[!t]
\centering
 \includegraphics[width=1.4in]{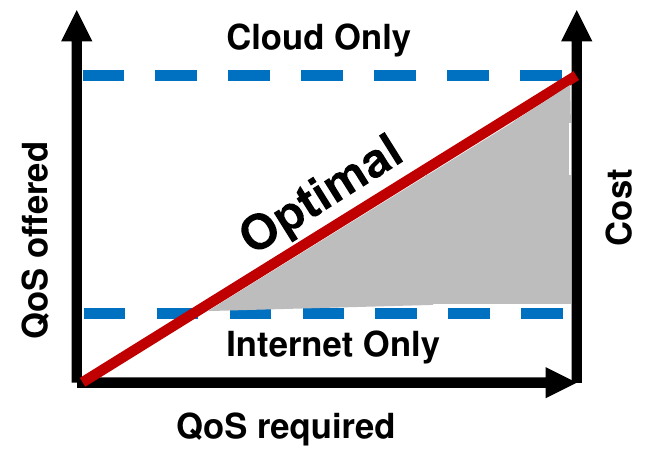}
 \caption{Judicious QoS.}
 \label{fig:solution-space}
\end{figure}

Figure~\ref{fig:solution-space} shows how this approach could be more efficient compared to 
today's Internet-only or cloud-only approaches. An Internet-only solution has low cost but offers low QoS as well. 
In contrast, a cloud-only solution offers superior QoS, but at a higher cost.
By judiciously using the cloud, we can approach the optimal line where
the QoS required equals the QoS offered.
The shaded area shows the cloud resources that are used
to meet the gap between the QoS required and the QoS offered by the Internet -- 
it is in-line with the spirit of cloud's pay-as-you-use model. In the next section, we show that it is indeed possible
to have cloud services that can allow such judicious use of cloud resources.

\section{\sys{} Design}
\label{sec:design}

\sys{} offers cloud-based reliability services that enhance the best effort service provided by IP.  
Figure~\ref{fig:services_overview} shows the main use-case for the \sys{} services: there is a sender (S) sending latency sensitive traffic (e.g., voice) to a receiver (R) over a wide-area Internet path (e.g., across continents). 
Both the sender and the receiver have nearby DCs (DC1 is close to the sender while DC2 is close to the receiver) with a small access latency ($\delta$) and the only cost incurred at the DCs is their egress bandwidth charge, which is denoted by $c$.   
These DCs run the \sys{} services, which leverage different aspects of the cloud  (storage, processing, etc), and offer trade-offs in terms of latency and cost, as we describe next. 

The \emph{forwarding} service forwards packets over the cloud overlay,  
leveraging the reliable inter-DC paths and high egress bandwidth of DCs. 
In the typical use case of forwarding, the sender forwards the packet to DC1, which forwards it to the receiver using the cloud overlay (via DC2), similar to VIA~\cite{via2016}. 
The resulting packet delivery latency is comparable to the direct Internet path latency, as we show in our evaluation (\S\ref{subsec:feasiblity}). However, it incurs the cloud bandwidth cost of $2c$ (egress bandwidth of both DC1 and DC2).

The \emph{caching} service, built on top of the forwarding services, provides on-demand delivery by storing packet at the DC.
In the typical case, a copy of the packet is sent on the cloud overlay -- from the sender to DC2 via DC1 --  but instead of forwarding it all the
way to the receiver, it is cached at the DC close to the receiver. 
In case of a packet loss on the Internet path, 
the receiver can initiate a pull request to get the missing packet
from the nearby DC (i.e., DC2). Compared to the forwarding scenario described above,
caching service can reduce the cost from $2c$ to $c$ but at the expense of additional delay which is at least $2\delta$.

Our \emph{coding} service (Figure~\ref{fig:coding_overview}) uses cloud processing and generates a small
number of coded packets at DC1; these coded packets are sent across the inter-DC path and cached at
DC2.  When a receiver tries to pull a missing packet, the 
DC undertakes a cooperative recovery process
with the help of other (nearby) receivers. 
This service is the extreme point in this design space --  it brings the cost down to only $\alpha \cdot c$ (where $\alpha$ is a small constant << 1). However, it adds some latency by requiring an additional delay $4\delta$ time once the receiver detects a loss. 
With latencies to nearby DCs ($\delta$) getting smaller  
and the best effort Internet being sufficient most of the time, the coding service
can be a cheaper (but with higher latency) alternative to the caching service, which in turn, is a cheaper alternative to the forwarding service.

\begin{figure}[!t]
\centering
  \includegraphics[width=2.5in]{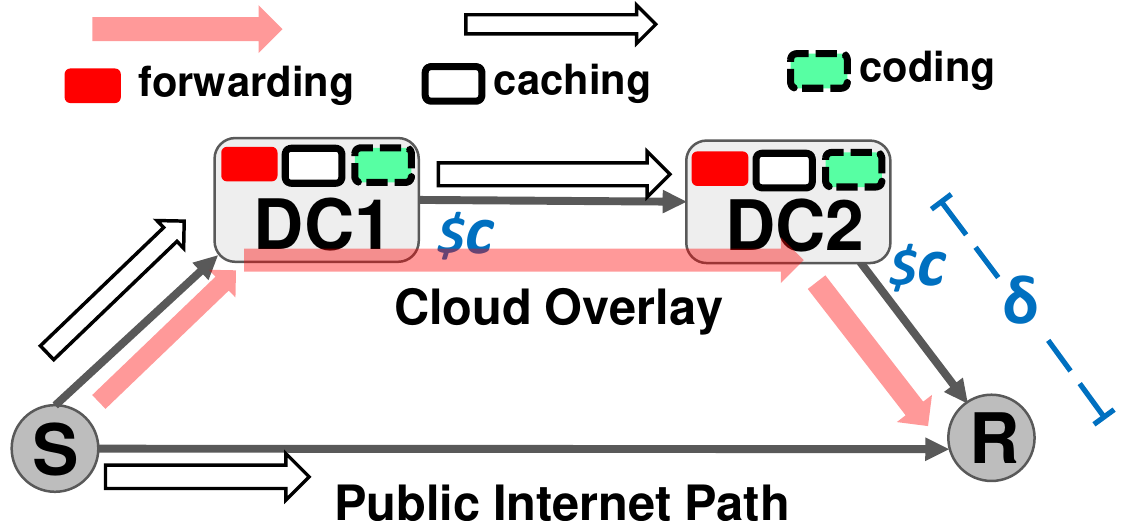}
  \caption{\sys{} Services - An Overview. Forwarding sends packets using the cloud overlay. Caching stores a copy of packet at DC2 while Coding only sends a small number of coded packets across the inter-DC path.}
  \label{fig:services_overview}
\end{figure}

To fully benefit from these services, \sys{} also provides suitable end-point support, in the form of a reliability layer that logically sits in between the transport and network layers, thereby enabling support for legacy applications, including both encrypted and non-encrypted traffic. 
Unlike today's sender-based recovery mechanisms (e.g., TCP), \sys{} includes
a receiver-driven recovery protocol (\S\ref{subsec:loss_detection_design}), which proactively detects losses, and undertakes loss recovery with the help of a nearby DC. 
Finally, \sys{} employs a simple API and service selection mechanism (\S\ref{subsec:api_framework}): given an application latency budget, it
chooses the cheapest service that can meet the requirement. 

The above simplified overview only considers the bandwidth cost for the services, but in reality, the cost depends on the deployment scenario. In \S\ref{subsec:deployment_model_cost}, we discuss how bandwidth costs would dwarf other costs (e.g., processing) if \sys{} services are deployed on a public cloud (e.g., Azure), and how advance services like coding would provide significant cost savings over a service like forwarding. 
Finally, we also discuss how \sys{} services can be used to support a diverse range of scenarios -- beyond the primary use case considered in this paper -- including support for multicast, partial cloud overlays (where only a single DC is used rather than a full overlay), selective duplication, and different cost models (\S\ref{subsec:other_usecase}).

\subsection{Coding Service}

\label{subsec:code-design}
\begin{figure}[!t]
\centering
 \includegraphics[width=2.5in]{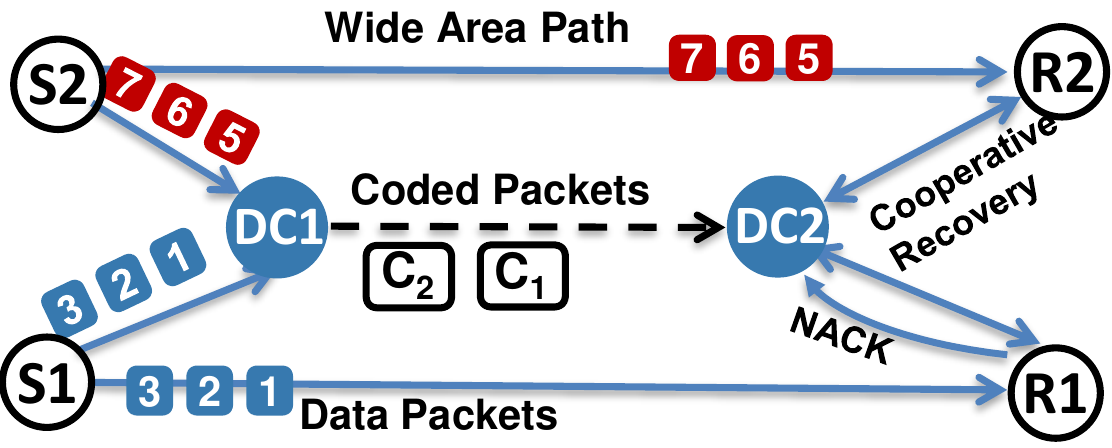}
 \caption{Coding Service Overview. DC1 encodes packets, DC2 stores them and performs recovery upon request.}
 \label{fig:coding_overview}
\end{figure}

\begin{figure*}[!ht]
\centering
  \hspace{-1em}
  \subfigure[\small cross-stream and
  in-stream encoding]{\includegraphics[width=1.13in]{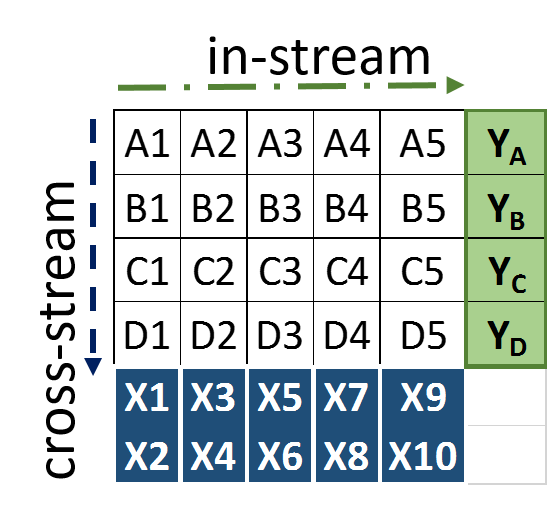}\label{fig:coding-1}}
\hspace{2em}
  \subfigure[\small $Y_A$~protects~flow 
  $A$ (in-stream).]{\includegraphics[width=1.0in]{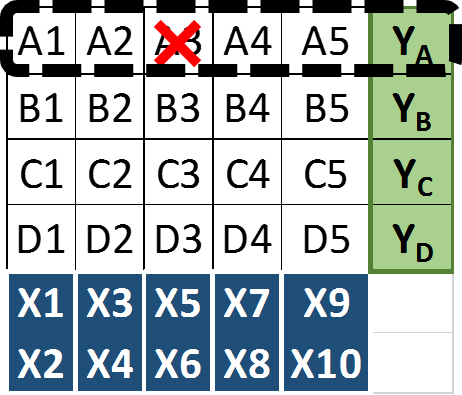}\label{fig:coding-2}}
 \hspace{2em}
  \subfigure[\small $X$'s protect flow $A$ (cross-stream).]{\includegraphics[width=1.0in]{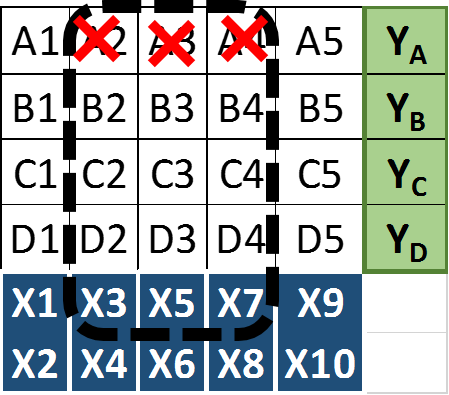}\label{fig:coding-3}}
\hspace{2em}
 \subfigure[\small $X$'s protect $A$
 and $C$ (cross-stream).]{\includegraphics[width=1.0in]{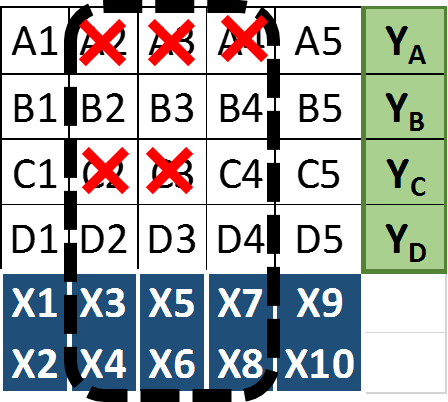}\label{fig:coding-4}}
 \hspace{2em}
  \subfigure[\small Cooperative Recovery.]{\includegraphics[scale=0.5]{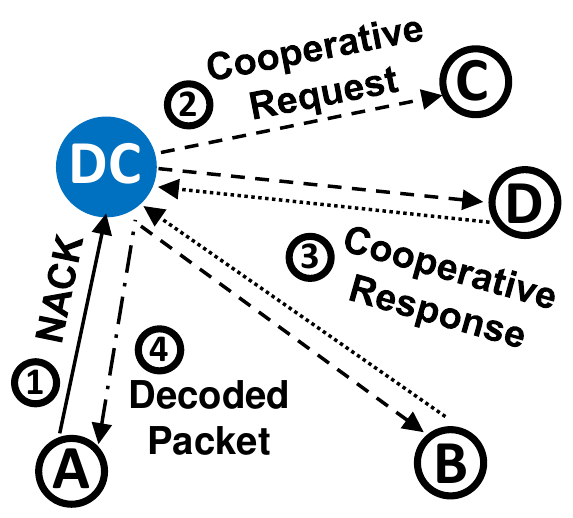} \label{fig:coop_recovery}}
  \caption{Coded Packet Generation and Recovery.}
 \end{figure*}

We focus on the design of the coding service as it makes use of both the forwarding and caching services. 
Our coding service, \rewan{}, uses a full overlay, where multiple senders send a copy of their packets to their nearby DC (Fig.~\ref{fig:coding_overview}). DC1 generates a small number of coded packets, which are sent to DC2 using the inter-DC cloud path. 
The key aspect of \rewan{} is in how it generates these coded packets -- some important considerations include which packets are considered together, what type of coded packets are generated, and at what rate.  
\rewan{} uses a novel \emph{cross-stream} coding design: coding is done \emph{across} a subset of user streams, which protects against bursty losses or even complete outages on a network path. For example, if ($S_1$-$R_1$) experiences an outage, \sys{} undertakes a \emph{cooperative} recovery process by combining the coded packets at DC2 with the data packets of $S_2$-$R_2$ to recover the lost packets. 

The cooperative recovery process, however, has its own set of challenges. 
First, decoding overhead can be high because it requires getting data packets from all other flows in the encoding subset. 
To ensure that this procedure is invoked only when necessary, \rewan{} also uses \emph{in-stream} coding,
whereby it generates a small number of forward error correction (FEC) packets \emph{within} a single user stream, thereby avoiding the potentially costly cooperative recovery 
for random losses.
Second, during cooperative recovery, some packets could be lost or delayed, especially if many streams are involved -- we call this the \emph{straggler problem}. \sys{}'s cross-stream coding accounts for potential stragglers by generating extra coded packets, thereby treating packets from stragglers similarly to losses on the direct Internet path. We elaborate on how \rewan{} deals with these challenges in \S\ref{subsec:coding_rate}.

While \rewan{} provides a cost-effective recovery option, its effectiveness depends on a number of factors, such as 
the latency and nature of losses on the direct Internet paths, latency between DC to end hosts, cloud's visibility into concurrent streams, and independent losses across multiple flows. Through \rewan{}'s deployment on PlanetLab (\S\ref{subsec:pl-eval}), we shed light on these factors and highlight scenarios
where \rewan{} is able to provide unique benefits compared to other potential loss mitigation techniques (e.g. FEC).

We now elaborate on the key pieces of \rewan{} encoding process: what packets are considered together for coding (the coding plan) and at what rate are the encoded packets generated (coding rate). 

\subsubsection{Coding Plan}
The coding plan needs to account for \emph{spatial} and \emph{temporal} constraints while forming a batch of packets on which coding will be applied. By spatial constraints, we mean that only flows with the same destination DC can be considered together for cross-stream coding. For example, if DC1 is in the East US region and is receiving traffic destined for a European DC and an Asian DC, it forms two groups, one for each destination DC.
Each flow belongs to one group and DC1 keeps a track of the mapping of flows to groups. 
Within a group, we pick a further subset of flows based on the arrival timing of their packets to form coding batches.

Temporal constraints restrict packets in a batch to only those packets that arrive within a short interval -- this imposes an encoding delay. For in-stream coding, the encoding delay is well-understood (and is considered a limitation of FEC for low bitrate applications) as we need to wait for all packets in a block to arrive before we can generate the FEC packets. However, \sys{}'s use of  cross-stream coding ensures that encoding delay is typically lower, because packets from different user streams can arrive within a short time-frame, even if each application individually is generating low bitrate traffic. Finally, our coding module limits the block size (for a given level of protection) and uses timeouts to bound delay. 

\subsubsection{Coding Rate}
\label{subsec:coding_rate}
Given a batch of data packets arriving at DC1, \sys{} needs to decide how many cross-stream and in-stream coded packets to generate. For both types, the coded packets are created using a block code (for example, Reed-Solomon codes), which allows \sys{} to generate multiple coded packets per batch if desired.  Figure~\ref{fig:coding-1} depicts some of the possible trade-offs, for a batch of 20 packets from four synchronous (for simplicity) flows, A-D.  In this depiction, in-stream encoding proceeds horizontally: a single FEC packet ($Y_i$) is produced for each flow $i$.  Cross-stream encoding proceeds vertically:  two cross-stream packets are produced from groups of four packets across flows, i.e., $A2$, $B2$, $C2$, and $D2$ are combined to generate coded packets $X3$ and $X4$.

Coding logically proceeds with two rates: an in-stream encoding rate of $s < 1$ coded packets per within-flow data packets, and a cross-stream encoding rate of 
$r < 1$ coded packets per data packet, where the data packets are selected among at most $k$ different flows.\footnote{We deviate from the standard notation of block coding theory, where $k$ data elements are encoded to generate a block of size $n$, yielding $(n - k)$ coded packets. Data rate and timing constraints may require us to code before $k$ packets are available.} 
Note that DC1 must also include information in the coded packets about which flows and sequence numbers are represented, to facilitate later recovery.
In our depicted setting, we have $k = 4$, $r = \frac{2}{4}$ and $s=\frac{1}{5}$, but in practice we use fewer coded packets for a batch of data packets, with the typical overhead of coded packets less than 20\%. 

Coded packets provide protection in multiple ways.  In-stream encoding packets protect primarily against random loss, much like traditional FEC, providing a first line of defense: providing faster recovery for random losses.
As depicted in Figure~\ref{fig:coding-2}, packet $Y_A$ can recover from the loss of $A_3$.
Cross-stream encoding, on the other hand, is both much more powerful (it can recover both random and bursty losses), but also incurs a potentially higher delay because of cooperative recovery(\S\ref{subsec:loss_detection_design}).
In Figure~\ref{fig:coding-4}, if some of $C$'s packets are also lost on the direct path, additional protection using more encoding packets could enable recovery at both $A$ and $C$.

\begin{algorithm}[!t]
\scriptsize
\SetKwFunction{KwFn}{dc1\_process(pkt, flow\_id):}
\textbf{def} \texttt{in\_stream\_qs[]}\\
\textbf{def} \texttt{cross\_stream\_qs[][]}\\
\BlankLine
\KwFn

\Indp
   \BlankLine
   \textit{// (1) In-stream coding.}\\
   \nl \texttt{q = in\_stream\_qs[flow\_id]}\\
   \nl \texttt{q.push(pkt)}\\
   \nl \If{\texttt{q.isFull()}} {
       \nl \texttt{in\_coded\_pkts = encode(q)}\\
       \nl \texttt{send(dc2\_id, in\_coded\_pkts)}\\
    }
    \BlankLine
    \textit{// (2) Cross-stream coding.}\\
    \nl \texttt{dc2\_id = extract\_dc2\_id(flow\_id)}\\
    \nl \texttt{q\_index = next\_round\_robin\_q(flow\_id)}\\
    \nl \texttt{q = cross\_stream\_qs[dc2\_id][q\_index]}\\
    \BlankLine
    \textit{// Find a queue that doesn't have a packet from this flow.}\\
    \nl \texttt{initial\_q = q}\\
    \nl \While {\texttt{q.contains(flow\_id)}} {
        \nl \texttt{q\_index = next\_round\_robin\_q(flow\_id)}\\
        \nl \texttt{q = cross\_stream\_qs[dc2\_id][q\_index]}\\
        \textit{// If we've tried all q's, empty the first by encoding or discarding.}\\
        \nl \If {\texttt{q $==$ initial\_q}} {
            \nl \If {\texttt{q.size() > 1}} {
                \nl \texttt{cross\_coded\_pkts = encode(q)}\\
                \nl \texttt{send(dc2\_id, cross\_coded\_pkts)}\\
            }
            \nl \Else {
                \nl \texttt{q.clear()}\\
            }
            \nl \texttt{\textbf{break}}\\
        }    
    }
    
    \BlankLine

    \nl \texttt{q.push(pkt)}\\
    \nl \If {\texttt{q.isFull()}} {
        \nl \texttt{cross\_coded\_pkts = encode(q)}\\
        \nl \texttt{send(dc2\_id, cross\_coded\_pkts)}\\
    }
 \caption{Coding algorithm at DC1.}
\label{alg:dc1}
\end{algorithm}
\subsubsection{Coding Algorithm.}
DC1 follows Algorithm~\ref{alg:dc1}, which captures the task of encoding across multiple flows at once. DC1 maintains two sets of queues: one set for in-stream encoding (one set per flow),
and a set for cross-stream encoding (one set per $k$).
When a packet arrives, it is copied and pushed into one queue of each type. 

Lines 1-5 check whether the relevant in-stream queue has reached a threshold, and if so, create coded packets and send them to DC2. For cross-stream coding, DC1 first selects the set of queues destined for the same DC2, and then chooses the individual queue in round-robin order (lines 6-8). DC1 avoids placing multiple packets from the same flow in the same cross-stream queue; if there already exist packets from the same flow in all queues, then DC1 processes the oldest queue. If there is only a packet from the flow in question, then the old packet is evicted and discarded, since sending cross-stream packets with only packets from a single stream reduces its effectiveness (lines 9-19). Once the packet is pushed into a cross-stream queue, if a threshold is reached, then coded packets are generated and sent to DC2 (lines 20-23).

Timing constraints pose a challenge to this algorithm. If one flow is much faster than all other flows, DC1 cannot hold back recovery data from the faster flow to wait to make full recovery packets. Therefore, we create a timer for each in-stream and cross-stream queue (not shown in Algorithm~\ref{alg:dc1}). On expiry of a queue timer, DC1 encodes all packets in the queue and sends them to DC2.

\subsection{Receiver-Driven Recovery Protocol}
\label{subsec:loss_detection_design}
For the caching and coding services, \sys{} uses a receiver-driven recovery protocol: the onus is on the receiver to quickly detect packet loss and undertake recovery with the help of its nearby DC. The key challenge in loss detection is how to make a fast, accurate prediction of whether a packet is lost (and thus needs to be recovered using the nearby DC). 
Note that our receiver based loss detection cannot use traditional sender-based timeout mechanisms (e.g., TCP RTO) because the receiver does not have a notion of \emph{when} a particular packet is sent by the sender. 

The caching and coding services provide (short term) storage of packets inside a DC. 
For any packet using these services, there should be an associated \emph{timeout} value and an \emph{identifier} that can be used to retrieve/pull that packet. These concepts are well known in the context of prior proposals that support in-network caching (e.g., NDN~\cite{Jacobson09}, XIA~\cite{xiansdi}, etc) or indirection-based architectures (e.g., i3~\cite{i3}).  
For the use cases considered in this paper, in-memory caching of packets is sufficient, 
but other scenarios could benefit from longer term storage of the packets (e.g., DTN~\cite{Fall03}, SlackStack~\cite{slackstack}). 
Similarly, any unique identifier schema (e.g., XIDs~\cite{xiansdi}, URIs~\cite{Jacobson09}, etc) can be used, although for convenience our prototype uses unique packet sequence numbers.

\paragraph{Loss Detection.} In \sys{}, the receiver detects a loss if
either a gap in sequence numbers is detected (the simple case) or a timer expires for the next expected packet. Setting a suitable timeout value -- low enough for fast recovery, but high enough not to cause spurious timeouts -- requires learning and predicting packet arrival times. While this opens up the possibilities to use machine learning algorithms, our current design uses a simple two-state Markov model that works well for our workloads. 
The model uses a \emph{small} timeout value for packets arriving within a burst (i.e., sub-RTT scale), and a \emph{long} timeout value across packet bursts or application sessions. These values are chosen based on previously observed inter-arrival times of packets and \sys{} loss detection module switches back and forth between them accordingly. 

Once a NACK is sent to the nearby DC, the recovery depends on the service being used. In the case of caching, recovery is simple as the data packet can be  transmitted to the requesting receiver. \rewan{}'s recovery could be more involved, as it may need to undertake \textit{cooperative} recovery.

\paragraph{Cooperative Recovery.} Figure~\ref{fig:coop_recovery} shows the steps involved in the cooperative recovery protocol.
After receiving a NACK from receiver $A$ (step 1), DC checks that there are sufficient cross-stream coded packets to conduct the recovery process. If so, DC 
then sends cooperative requests to 
relevant receivers since they have the data packets needed to decode the missing packets (step 2). The \sys{} module at the receivers stores the data packets for few(1-2) RTTs and purges them after this time.

DC then processes any incoming cooperative recovery responses from the solicited receivers (step 3).

By tracking responses, DC can tabulate the number of cooperative responses for each recovery event. 
For each loss, once the number of responses is equal to $k - 1$ 
then recovery is possible. DC then decodes the lost packets and sends them to the receiver (4). Depending on the number of cross-stream coded packets, DC may only require a few of the receivers to respond in a timely fashion, thereby ignoring stragglers (such as $C$ in Figure~\ref{fig:coop_recovery}) that can cause delay in recovery.

Since recovery is time sensitive, the protocol fails silently if not enough coded packets or cooperative recovery responses are received within a set deadline. We discuss these conditions under which recovery is not possible in Section~\ref{subsec:pl-eval}.

\begin{figure*}[!t]
\centering
  \subfigure[Multipath Routing]{\includegraphics[scale=0.75]{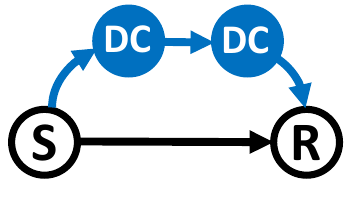}\label{fig:fwd_1}}
  \hspace{0.5em}
  \subfigure[Partial Overlay Routing]{\includegraphics[scale=0.75]{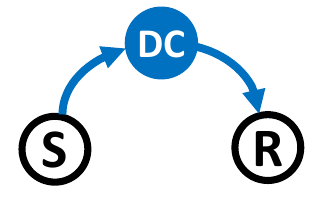}\label{fig:fwd_2}}
  \hspace{0.5em}
  \subfigure[Multicast]{\includegraphics[scale=0.75]{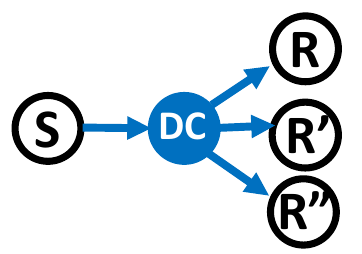}\label{fig:fwd_3}}
  \hspace{0.5em}
  \subfigure[Hybrid Multicast]{\includegraphics[scale=0.75]{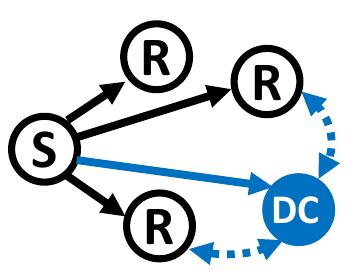}\label{fig:cache_1}}
  \hspace{0.5em}
 \subfigure[Mobility]{\includegraphics[scale=0.75]{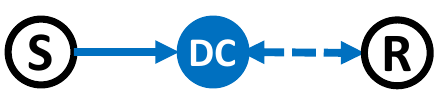}\label{fig:cache_2}}
\caption{Use cases for forwarding and caching service.}

\label{fig:overview_design}
 \end{figure*}

\subsection{\sys{} End-to-End Workflow}
\label{subsec:api_framework}
\sys{} services not only require adequate end-point support in the form of loss detection and recovery but also a way for applications to specify their demands.
We now describe how \sys{} framework selects and switches between services according to application demands and changing network conditions.

\paragraph{API and Service Selection.} Applications registering with \sys{} use a simple \texttt{register(...)} API to express their latency budget and target destination. Based on the 
budget, \sys{} selects the lowest cost service that can meet this requirement. 
As our services operate on a continuous spectrum, with \rewan{} being the cheapest and forwarding being the expensive, 
\sys{} service selection module picks the cheapest service as long as it can meet the latency budget. 
In our evaluation, we show that these services operate in different delay regions and can be  mapped to a latency demand (\S~\ref{subsec:feasiblity}).

End hosts using \sys{} also use \texttt{register(...)} to bootstrap with the membership management module running on a central DC. This module relays nearest DC location as well as latency information, required for service delay computation, to the registering end host.

For packet routing, our base forwarding service decides the next hop based on the destination address of the packet. 
Given the small scale of the overlay network in \sys{}, the next hop decision is simple and made in a centralized fashion. The next hop could be another \sys{} service, an end-point (e.g., the receiver), or a multicast group.

\paragraph{Delay Computation.}  \sys{} service selection module uses the destination of the application flow to calculate the service delay.
Some of the delays, such as latency between \texttt{DC1-DC2} are pre-computed and stored at each end host during initial bootstrap. Other delays such as \texttt{S/R-DC} latency ($\delta$) and \texttt{S-R} latency are initially assumed to be average values based on the latency of existing end-points communicating with their nearby DCs. The delay values are updated once communication starts between the end-points.

\paragraph{Feedback.} Finally, the service selection decision is communicated to the \sys{} sending module so it can route the packets accordingly. 
The service selection mechanism receives the packet delivery statistics from the receiver and can decide to upgrade to an improved service, if the existing service is not meeting the applications latency demand. As our caching and coding services mask packet loss from the sender, \sys{} can use this feedback information to indicate congestion to higher layers that may require this information (e.g. TCP).

\subsection{Deployment Model and Cost}
\label{subsec:deployment_model_cost}

\paragraph {Deployment model.} We believe a \sys{}-like service can potentially be deployed by the cloud infrastructure provider, an ISP, or a third-party service provider who pays the cloud operator only for the infrastructure usage. Users can explicitly opt into using \sys{} if they require a more reliable packet delivery service -- for example, implemented as a paid service or added value proposition. One recent example of a similar paid service is the network tier service by Google Cloud~\cite{gcloud}, which offers packet delivery using either standard transit ISP (at lower cost) or Google's own network (at higher cost) between its DCs~\cite{gcp-tiers}.

The deployment model may also have implications on the practical usage of \sys{} design. For example, an ISP or cloud provider with full control over its resources (e.g., using SDN~\cite{swan}) may prioritize the inter-DC traffic for even higher resilience and lower delay, leading to even higher recovery efficiency. 
An ISP providing the service can also reduce sender overhead by duplicating the packets in the core as well.

\paragraph{Deployment Cost.}
For estimating deployment cost, we consider the deployment scenario where a third-party provider wants to deploy \sys{} services over a public cloud. 
For a typical video conferencing scenario, We do a back-of-the-envelope calculation that compares coding in \sys{} with a solution that fully uses the cloud, such as the forwarding service.

Based on the bandwidth requirement of HD Skype call (1.5Mbps~\cite{skype_bw}), a single user will send 0.675 GB of data per hour. Given that  a single user space \sys{} encoding thread in our prototype can handle 150 Skype calls, a data center node will receive and forward $\sim$101 GB of data per hour, for 150 parallel application sessions. For a 2-node overlay node with forwarding, based on today's cloud pricing~\cite{azureprice}, this would cost a minimum of \$17.60/hour for bandwidth and only \$0.13/hour for single thread general purpose compute usage. However, 
for a coding rate of $r=1/15$, 
the \emph{maximum} cost of bandwidth for 150 calls will only be \$1.17/hour, which is 15x less than the cost of forwarding. Note that there is no storage cost because packets are cached temporarily in the local (free) storage of the VMs. 
In this calculation, we are assuming that every coded packet will be used to recover a lost packet, which is an upper-bound --  in practice, the outgoing bandwidth from DC2 will only be used in case of a packet loss.

\subsection{Other Use Cases}
\label{subsec:other_usecase}

We now discuss a number of other use-cases, involving the services (forwarding and caching), how applications use these services (e.g., selective duplication), and cost models. 

\paragraph{Forwarding.}
For even higher reliability, we can use \emph{both} the Internet and the cloud overlay: the sender
can use the forwarding service to transmit a copy of the packet to the receiver (Fig.~\ref{fig:fwd_1}). This use case is most beneficial for mission critical applications like financial transactions~\cite{citrix-dup}.
In contrast, a \emph{partial overlay} (single DC) scenario may not offer the benefits of reliable inter-DC
paths, but it costs less, and can still benefit from the high egress bandwidth of the DC (Fig.~\ref{fig:fwd_2}). 
This use case can be extended to support a multicast scenario (Fig.~\ref{fig:fwd_3}): the sender sends its stream to the cloud forwarding
service which forwards it to the multicast group, again leveraging its high egress bandwidth. This scenario can be useful for applications like video streaming, software distribution, etc.

\paragraph{Caching.}
The basic caching use-case described earlier could be extended to support a \emph{hybrid multicast}, which provides a cheaper alternate to the cloud-based multicast that uses the forwarding service only. In hybrid multicast (Fig.~\ref{fig:cache_1}), the sender uses the public Internet
to send its stream to all the receivers; a copy of the stream is sent to the nearby DC where it is cached. 
If a receiver fails to receive a packet, it goes to the DC and retrieves it.

The caching service is also useful for mobility scenarios, 
providing an on-path caching/rendezvous point for mobile-hosts, similar to 
Internet architectural proposals like NDN~\cite{Jacobson09}, XIA~\cite{xiansdi} and i3~\cite{i3}. 
For example, a mobile sender only sends packets to a DC, where they are cached.
The receiver, whenever it is online, pulls the packets from the nearby DC 
rather than
requiring the mobile sender to come online and retransmit the packets (Fig.~\ref{fig:cache_2}). 

\paragraph{Selective Duplication.} \sys{} can also support scenarios where applications may want only some packets to be duplicated and sent through the \sys{} services. Some examples of such packets could be an  I-frame for video streaming, important user actions for gaming or AR applications, and the last packet of a window for short TCP transfers~\cite{gentleaggression}. Such selective duplication can provide a lower overhead alternative to the full duplication option described earlier, and can be used in scenarios where sources have limited bandwidth or limited budget for packet recovery. We evaluate benefits of selective duplication in \S\ref{subsec:rewan-tcp-eval}.

\paragraph{Other Cost Models.} Our cost estimation assumed that ingress bandwidth is free, as is the case in current pricing schemes~\cite{amazonprice,azureprice,googleprice}. Even if ingress bandwidth is charged, the coding and caching services will still be more cost efficient compared to the simple forwarding service. Similarly, we considered the same egress bandwidth cost for different DCs, even though in practice, some regions (e.g., Asia or South America) may have higher costs. This has implications on the cost benefits of coding and caching services over the simple forwarding service. For the caching service, if the expensive DC is close to the receiver, it will provide even more gains compared to if it is close to the sender. The coding service, in contrast, reduces egress bandwidth usage on both DCs, so its gains are less sensitive to the placement of the expensive DC.

\section{\sys{} Prototype}
\label{sec:prototype}

The \sys{} prototype\footnote{Available at \url{https://tinyurl.com/jqos-code}} is implemented in C++ and operates in user space. 
Our implementation uses UDP for forwarding application traffic, coded packets, and cooperative recovery packets, and uses TCP for control channel traffic between the endpoints and the data centers. Applications can utilize \sys{} in two ways. First, iptables~\cite{iptables} and the NetFilter library~\cite{netfilter} can be used to install forwarding rules designed to ``catch'' outbound application traffic, redirect it to \sys{} to duplicate or forward it on the data center path.
Alternatively, \sys{} can act as a proxy listening on a local port for applications to directly send data. The data is received and encapsulated in the \sys{} header before being sent to the destination and DC1.

Our prototype uses Reed-Solomon codes to encode and decode application data using the open-source zfec~\cite{zfec} library, and uses 
25ms for the small timer and RTT for the long timer.
Finally, we tune the parameters related to coding (coding rate, timers, and queues) on a per-application basis, depending on the application's characteristics and requirements.

\paragraph{Coding Parameters.} For cross-stream coding, we use a default of two cross-stream coded packets ($r = 2/k$) to mitigate the effects of stragglers and protect against bursty losses and outages.
In practice, we bound $k$ to a moderate value ($k <= 10$ in our evaluation), since larger values add significant overhead in the cooperative recovery process. When more than $k$ flows use \sys{} concurrently at an ingress DC, the DC organizes them into subgroups of at most $k$ flows per group.

For in-stream coding, we find that for interactive applications -- where the average frame rate is 10-15 fps and the average frame is composed of 2-5 packets~\cite{chitchat} -- it is best suited to send an in-stream packet for each frame ($s=\frac{1}{5}$), although that results in relatively higher overhead, so applications with low cost budget can choose to fall back to cross-stream coding only. The in-stream encoding overhead is less for applications that send back-to-back packets, such as 
TCP flows, where a single coded packet can be sent for an entire TCP window (e.g., $s=\frac{1}{16}$ or $s=\frac{1}{32}$).

\section{Evaluation}
\label{sec:evaluation}

\begin{figure*}[!t]
\centering
\hspace{-0.7em}
  \subfigure[]{\includegraphics[width=1.8in]{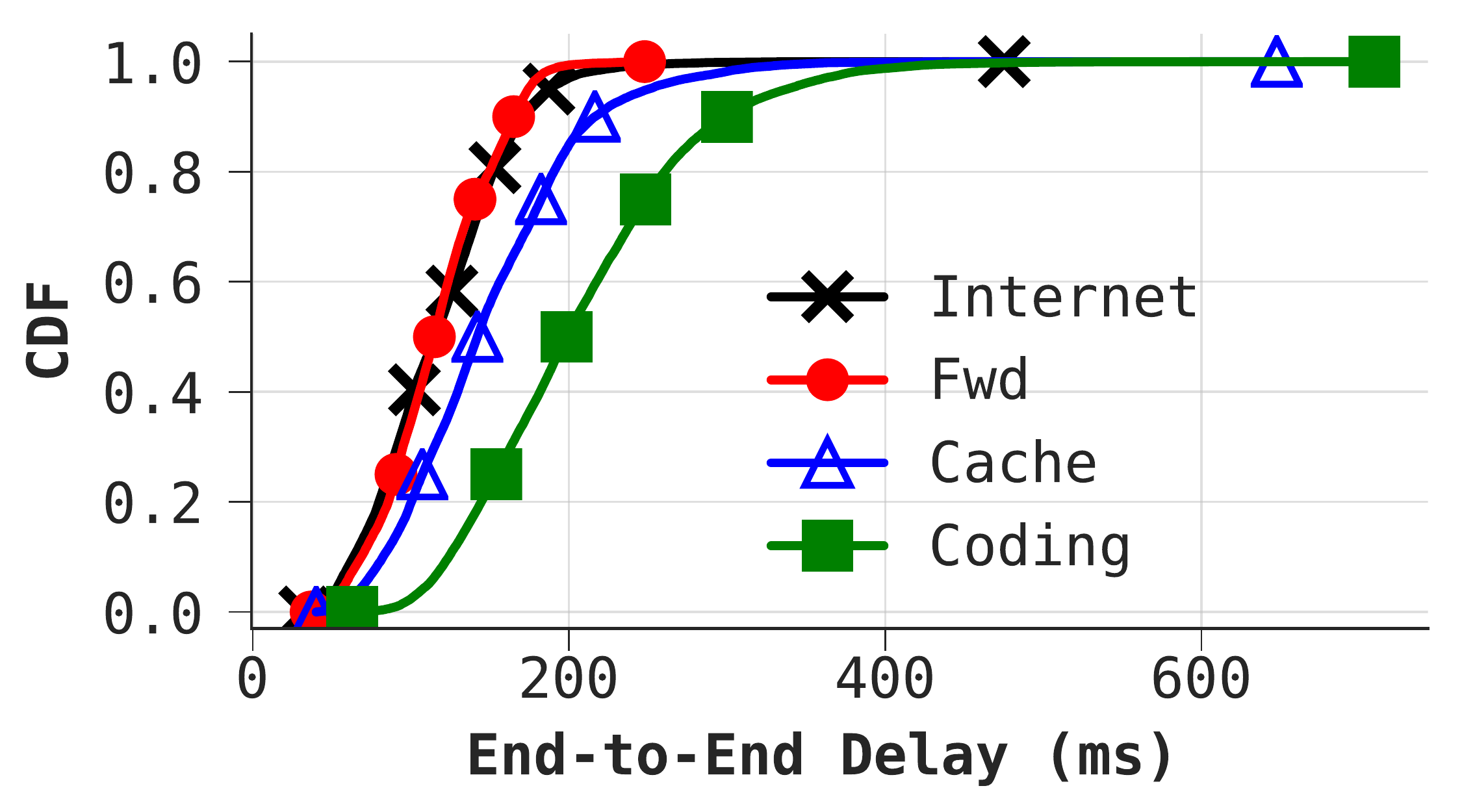}\label{fig:service_raw}}
  \hspace{-0.9em}
  \subfigure[]{\includegraphics[width=1.37in]{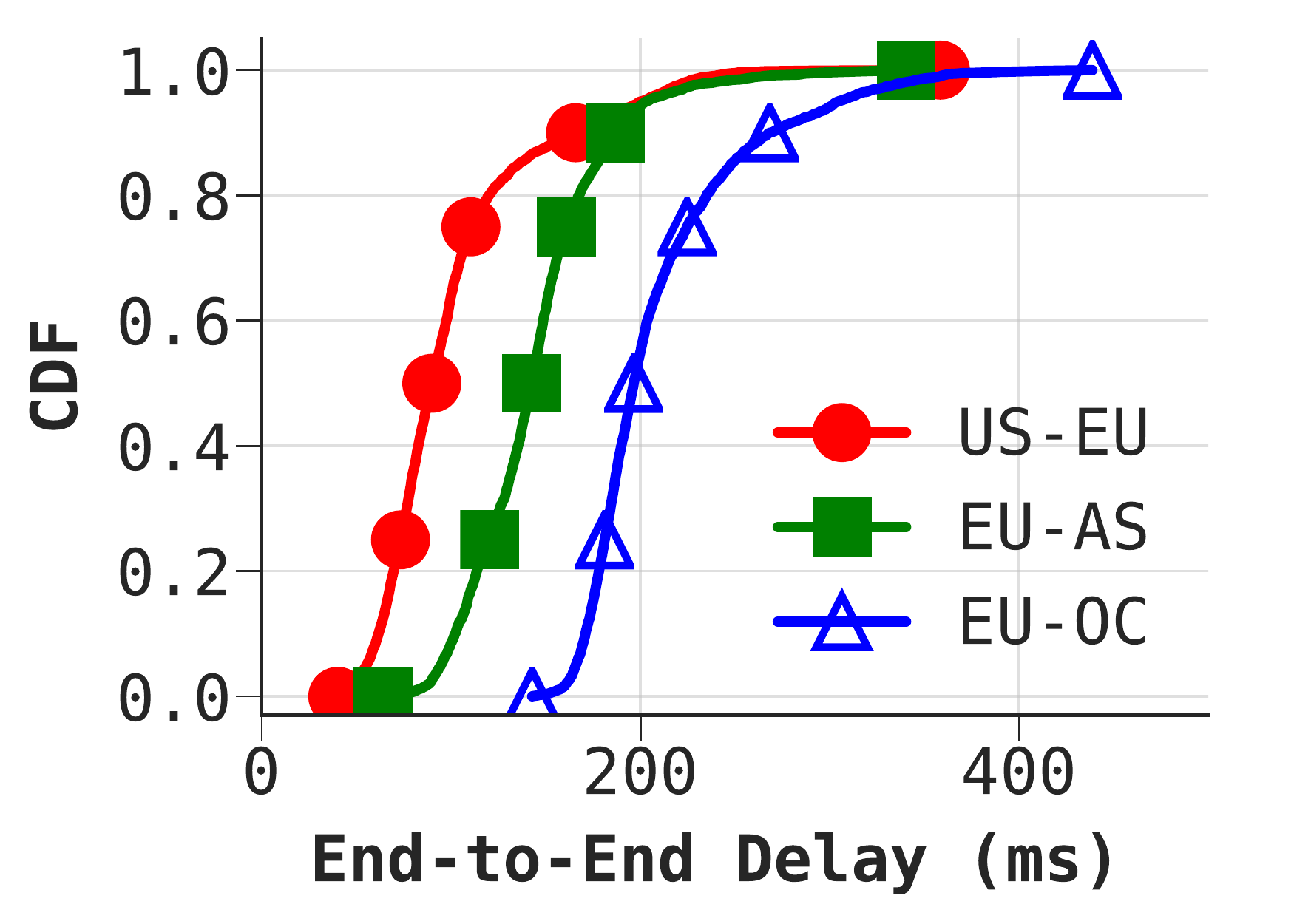}\label{fig:per_region_latency}}
  \hspace{-0.8em}
  \subfigure[]{\includegraphics[width=1.37in]{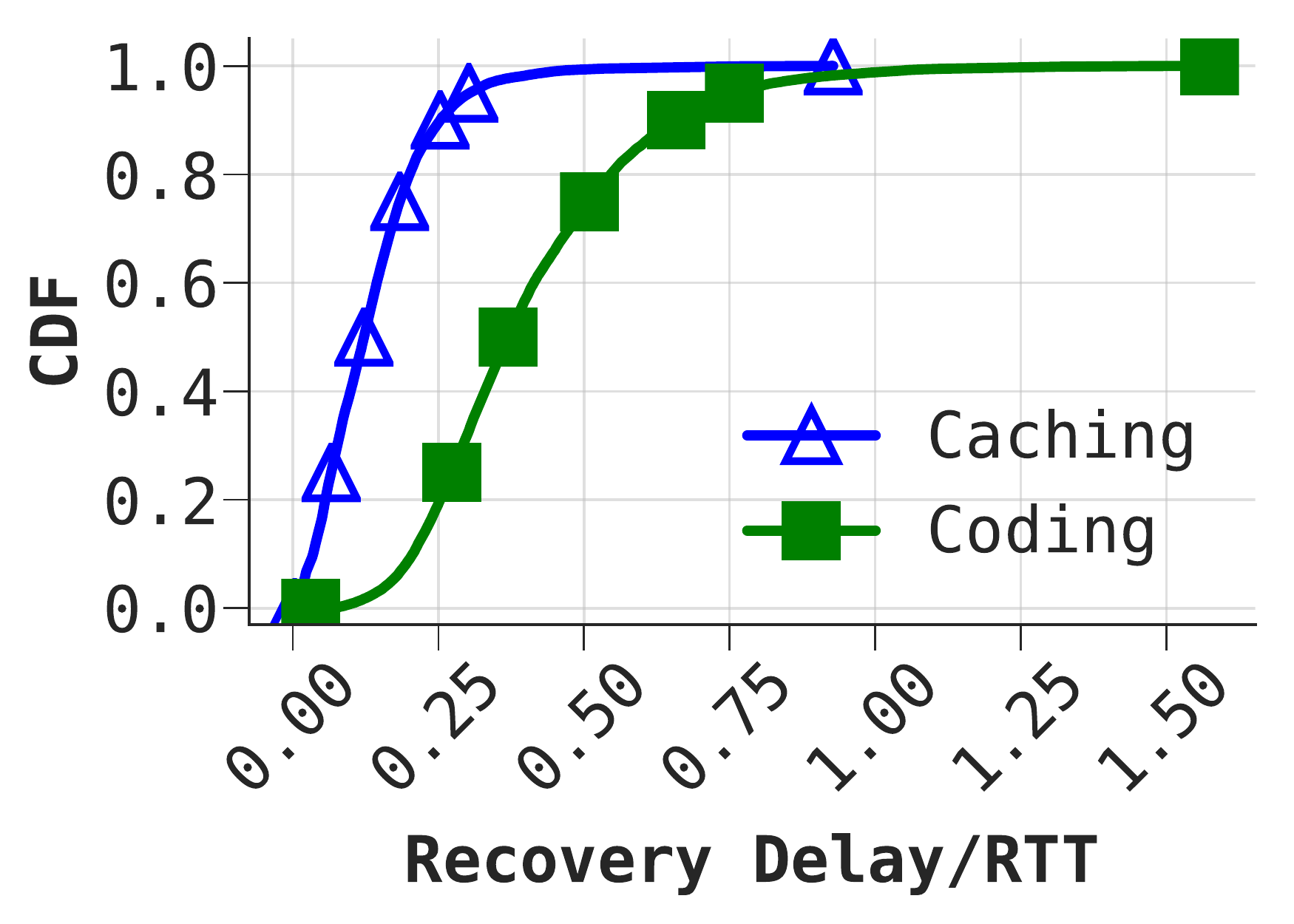}\label{fig:overlay_inflation}}
  \hspace{-0.8em}
  \subfigure[]{\includegraphics[width=1.37in]{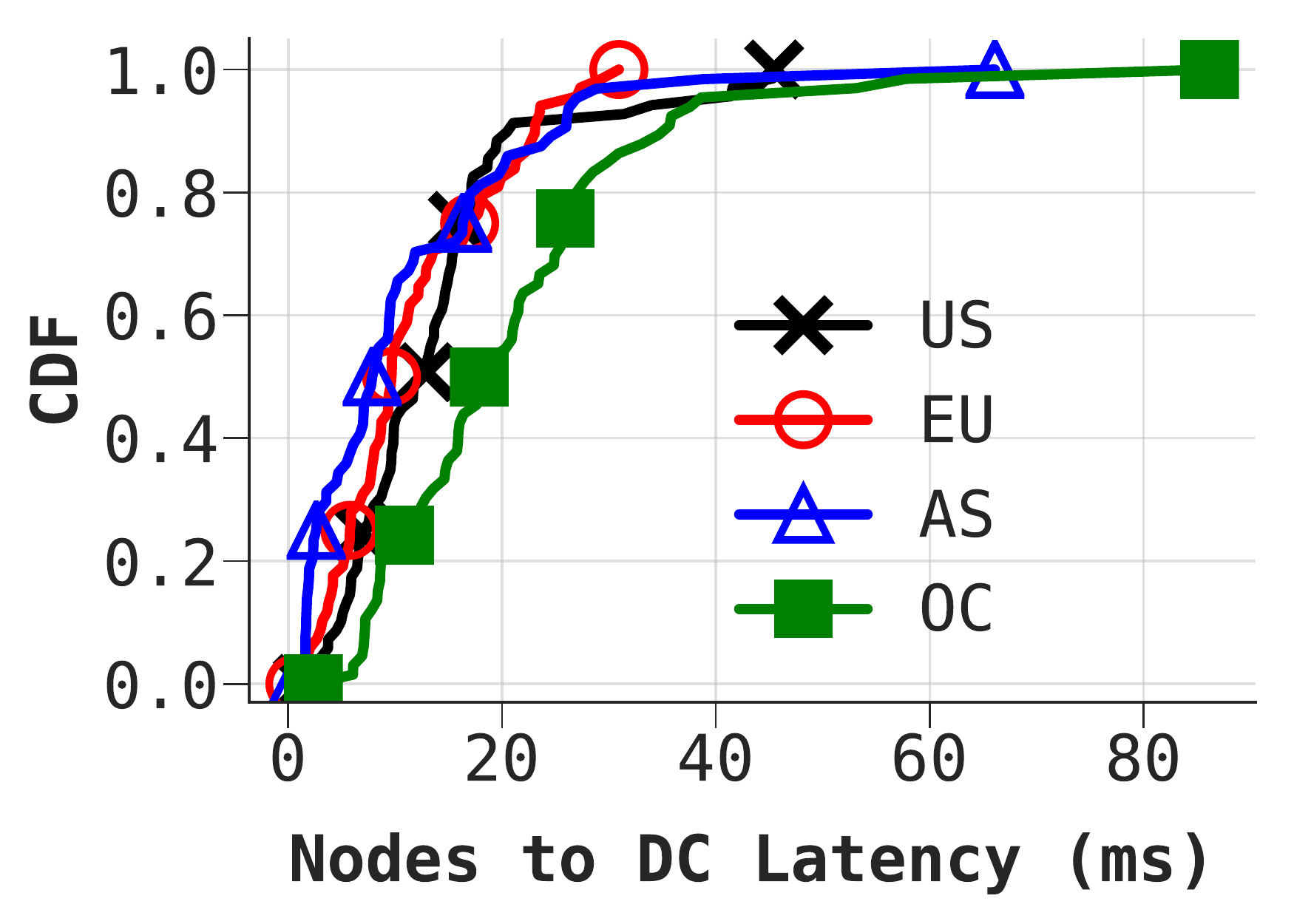}\label{fig:rtt_nearest_dc}}
  \hspace{-0.8em}
  \subfigure[]{\includegraphics[width=1.37in]{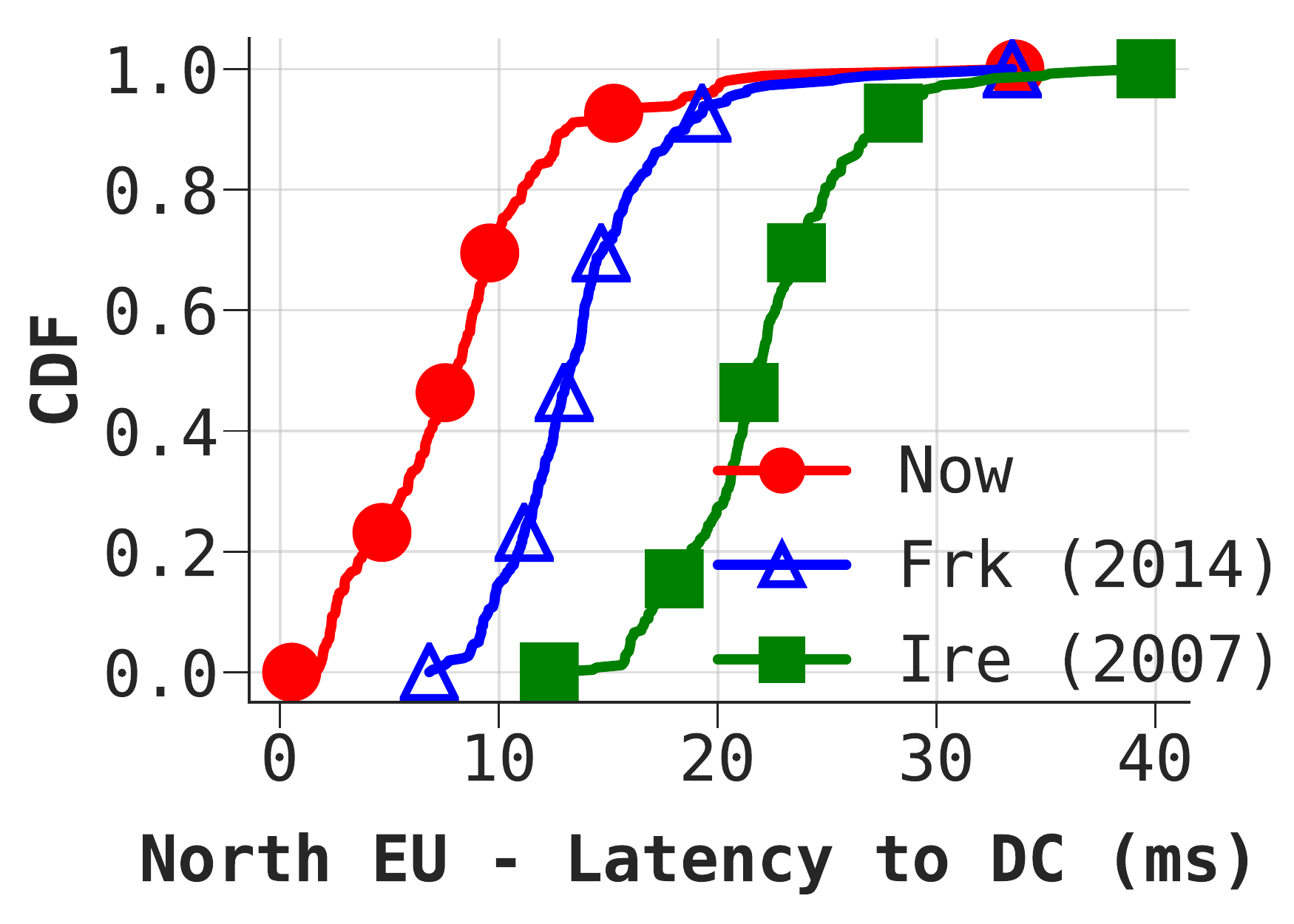}\label{fig:decreasing_dc_rtt}}
  \caption{\sys{} service feasibility. a) End-to-End packet delivery latency is within acceptable range. b) Caching service latency varies across regions. c) Most  packets are recovered within 0.5xRTT. d) Latency to nearest DC is small. e) Improvement in Latency to nearest DC over the years for northern EU hosts.}
\label{fig:ripe_dc_study}
 \end{figure*}

We perform a multi-tiered evaluation with the goals of answering: 
 (1) How feasible are \sys{} services in real world (\S\ref{subsec:feasiblity})?
 (2) How effectively does \sys{} coding service recovers packets within a time budget for wide-area paths (\S\ref{subsec:pl-eval})?
 (3) How does \sys{} perform in the contexts of challenging application (\S\ref{subsec:skype-perf}), transport (\S\ref{subsec:rewan-tcp-eval}), and network requirements (\S\ref{subsec:mobile-eval})?

\noindent Our evaluation encompasses:

\noindent\textbf{Feasibility Study on RIPE Atlas.} 
We perform latency measurements on 22k paths across the globe and compute \sys{} service delays(\S~\ref{subsec:feasiblity}). We find that forwarding service does not inflate latency whereas caching service can deliver packets within 200ms for 85\% of paths (50\% for coding). 

\noindent\textbf{Wide-area Deployment and Evaluation.} 
We ran \rewan{} on a public cloud for over a month and measured its effectiveness in recovering losses for 45 inter-continental PlanetLab paths (\S\ref{subsec:pl-eval}). Our results show that \sys{} is able to recover 78\% of the packet losses. Further, in  over 80\% of cases, the recovery time is less than half an RTT of the direct path between the sender and receiver. 
     
\noindent\textbf{Cross-Layer, Controlled Environment Evaluations.}
 We evaluate \sys{} with respect to user-level QoE metrics for Skype video conferencing (\S\ref{subsec:skype-perf}), flow completion time for short TCP flows (\S\ref{subsec:rewan-tcp-eval}), and network QoS in a cellular setting (\S\ref{subsec:mobile-eval}). We find that in controlled environments, \sys{} can provide benefits up and down the stack, albeit with some environment-specific tuning to keep it cost-effective.

\subsection{Feasibility of \sys{} Services}
\label{subsec:feasiblity}
Our goal is to evaluate the feasibility of \sys{} services using latency data from hosts around the world.

\paragraph{Methodology.} We use the RIPE Atlas testbed~\cite{ripe-atlas} and Amazon Web Services (AWS)~\cite{aws} data centers to measure latencies of public Internet and cloud overlay paths. In our scenario, RIPE Atlas anchor nodes are senders and probe nodes are receivers. We use AWS data centers near senders and receivers to form a full (2-DC) cloud overlay. Overall, we measure 22K paths spanning four continents, US, EU, Asia, and Oceania\footnote{Measurement details at: \url{https://tinyurl.com/atlas-paths}}.

We measure the following latencies for each pair of RIPE Atlas nodes:  $\delta$ (\texttt{S/R-DC}), $x$ (\texttt{DC1-DC2}), and $y$ (\texttt{S-R}). 
We compute delay for forwarding service as $x + \delta_{S-DC1} + \delta_{R-DC2}$, caching as $y + 2\delta_{R-DC2} + \Delta$ and coding as $y + 2\delta_{R-DC2} + 2\delta_{{R'-DC2}} + \Delta$. Note that, for caching and coding, we include $y$ which is one way latency from $S$ to $R$. We use $\Delta$ to represent the delay of a caching/coding pull request, if it reaches DC2 before the desired packet arrives from the sender to DC2. We also represent cooperative recovery delay as $\delta_{R'-DC2}$.  In this case, we compute it as the maximum \texttt{R-DC2} latency of five random nodes in the same region.

\paragraph{\sys{} services can meet latency budget of applications.} 
Figure~\ref{fig:service_raw} shows the end-to-end packet latency for \sys{} services as well as the direct Internet latency for all paths. 
We make three observations. 
First, using the (indirect) cloud overlay does not inflate latency compared to using the direct Internet: for a majority of the paths, 
the forwarding service has a latency similar to the Internet paths. 
Second, Internet delivery has a long tail compared to the forwarding service, confirming
earlier findings by Microsoft~\cite{via2016} that some Internet paths are persistently low quality,
and it is better to completely switch to a cloud overlay for such paths. The forwarding service
can be a great fit for such cases.
Third, we also observe that packet delivery, using the caching service, takes up to 200ms for 85\% of the paths (50\% for coding). This is acceptable delay for many latency-sensitive applications that require timely packet delivery~\cite{itu-qos}.

These results add to the growing evidence that cloud overlays can be feasible for a diverse range of end-to-end scenarios (e.g., cloud middleboxes~\cite{middleboxsherry}, web transfers~\cite{single-hotcloud}), etc).

\paragraph{Service delays vary across regions.}
Figure~\ref{fig:per_region_latency} shows the region-wise end-to-end packet delivery latency for the caching service. 
We observe that, in the US-EU region, packet delivery takes less than 200ms for 95\% of the paths.
We ascribe this to the short one-way latency between the majority of the nodes in the US and EU regions as well as small delta in the EU region. It also means that, for US-EU paths, caching and coding can provide similar benefits to the applications as the forwarding service. 
On the other hand,  due to high one-way latency on EU-OC paths, only 50\% of the paths deliver packets within 200ms, requiring the rest of the paths to rely only on forwarding. We also find that other paths (not shown e.g.US-AS, US-OC) follow a similar pattern as EU-AS paths.
For these paths,  
stricter delay requirements will result in a subset of the paths using caching service.
Region-wise coding service delays (not depicted) follow a similar pattern, albeit shifted slightly to the right.

These results show that based on the delay demand and the end hosts region, we can potentially select a suitable service.

\paragraph{On-demand Recovery Delays.}
A traditional retransmission-based recovery from the source takes at least one RTT whereas
\sys{}'s caching and coding services retrieve the missing packet from a nearby DC. 
We compare this difference 
by plotting the recovery latency as fraction of the RTT. 

Figure~\ref{fig:overlay_inflation} shows that 75\% of the time, these services can recover packets within 0.5xRTT of the direct Internet path. We also observe a clear separation between service recovery times, e.g., 90\% of the time, caching service can recover packets within 0.25xRTT whereas coding service can only recover packets 20\% of the time. Most of the coding service benefits are in 0.25xRTT to 0.5xRTT range, we also confirm this in our real world deployment (\S\ref{subsec:pl-eval}).

\paragraph{End host to DC latency ($\delta$) is small.} 
\sys{} services rely on end hosts having low latency to their nearest DC, so we now focus on these latencies. 
Figure~\ref{fig:rtt_nearest_dc} shows the $\delta$ value for RIPE Atlas probes in different regions (i.e., EU, Asia, Oceania, US). 
We observe in  that 50\% of paths have $\delta$ less than 10ms in Europe and Asia. We also observe that 20\% of paths have delta higher than 20ms. While Oceania region has relatively high $\delta$ compared to the other regions, we observe that 75\% of probes can be reached within 25ms. Depending on the application's latency budget, paths with high $\delta$ can still utilize coding and caching services.

\paragraph{Delays for services will decrease in the future ($\delta$ is becoming smaller).} In Fig.~\ref{fig:decreasing_dc_rtt}, we focus on the receivers in northern EU and evaluate how the emergence of new DCs in Europe may have impacted their latency to their nearest DC. The ``Now'' result shows the latency of these nodes to the Stockholm DC, which was opened in 2018. Before that the nearest DC for these nodes was the Frankfurt DC (opened in 2014) and even before it was Ireland (opened in 2007). We observe that $\delta$ is decreasing over time and nodes all over the world could potentially 
see similar improvement in future. 

\begin{figure*}[!t]
\centering
  \subfigure[]{\includegraphics[width=1.37in]{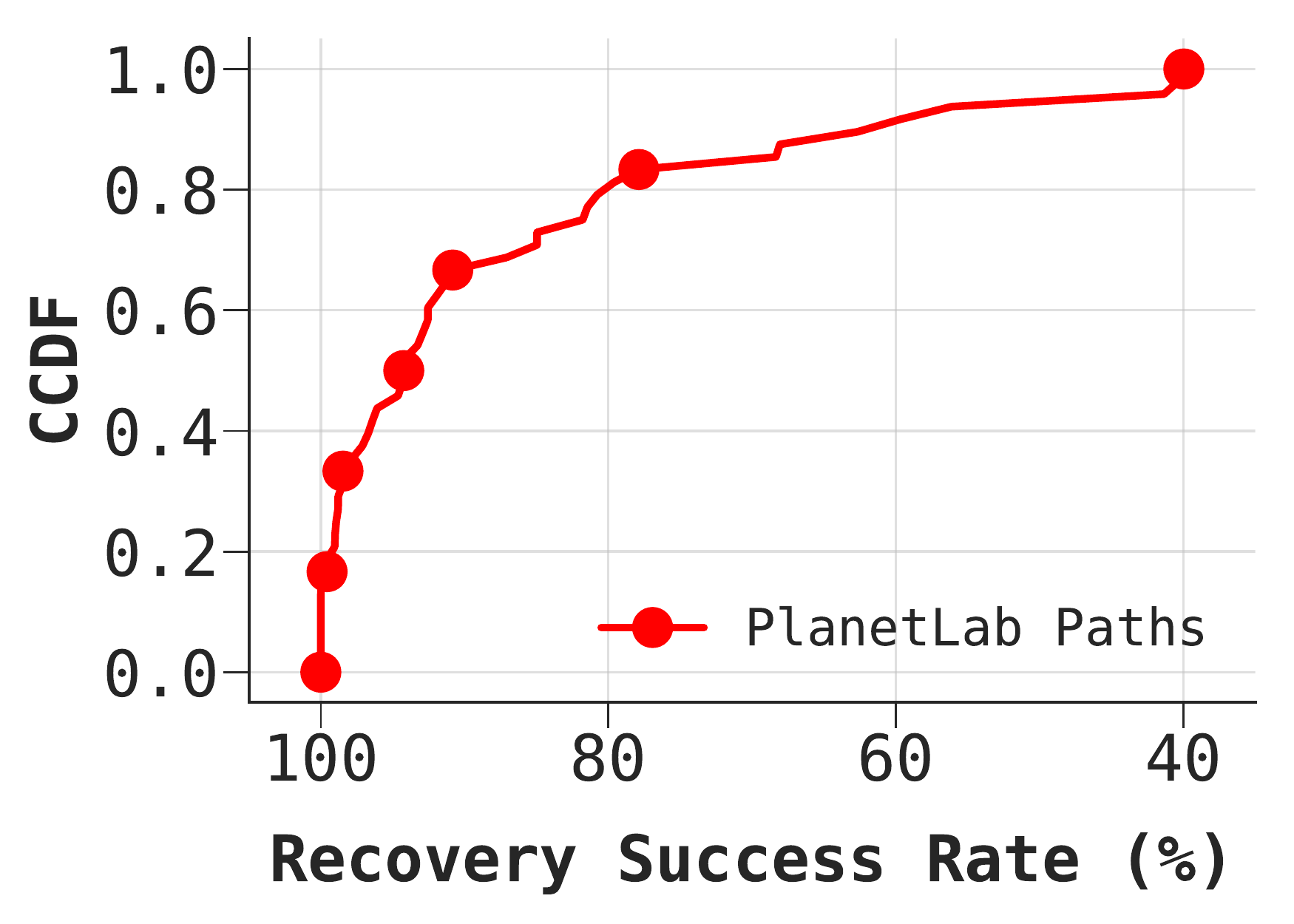}\label{fig:rewan_recovery_overview}}
  \subfigure[]{\includegraphics[width=1.37in]{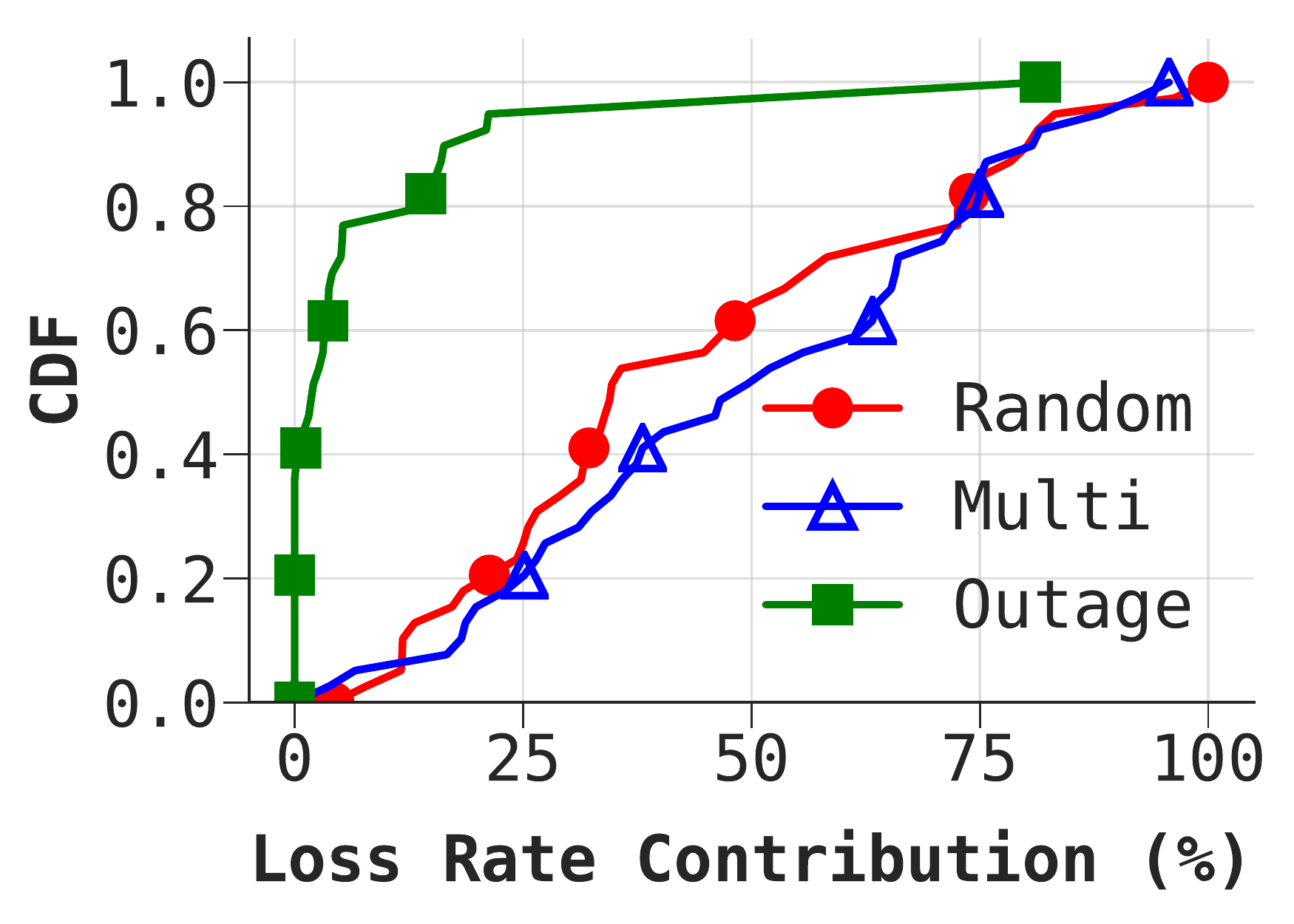}\label{fig:rewan_burst_overview}}
  \hspace{0.01em}
  \subfigure[]{\includegraphics[width=1.37in]{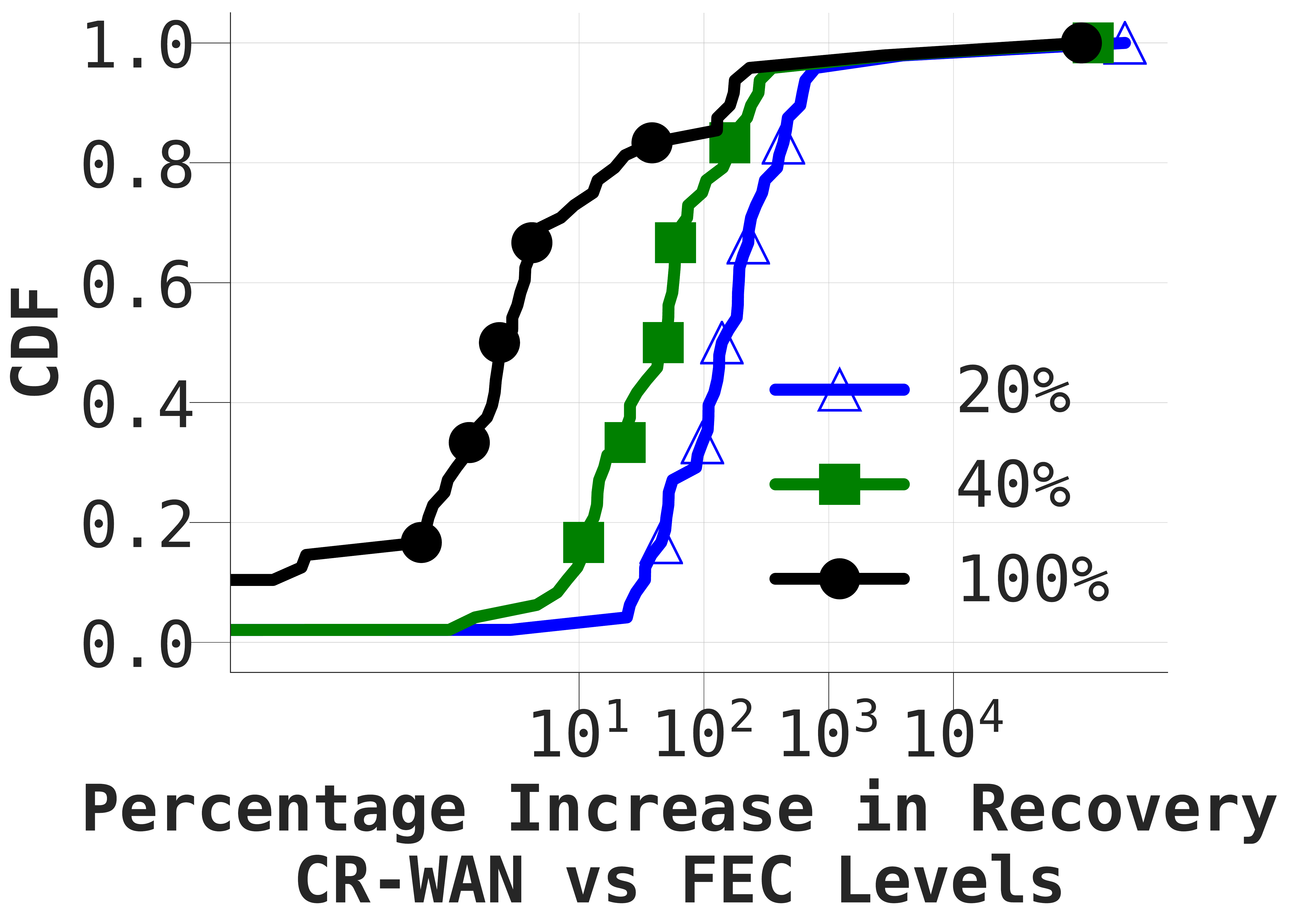}\label{fig:rewan_fec_diff}}
  \subfigure[]{\includegraphics[width=1.37in]{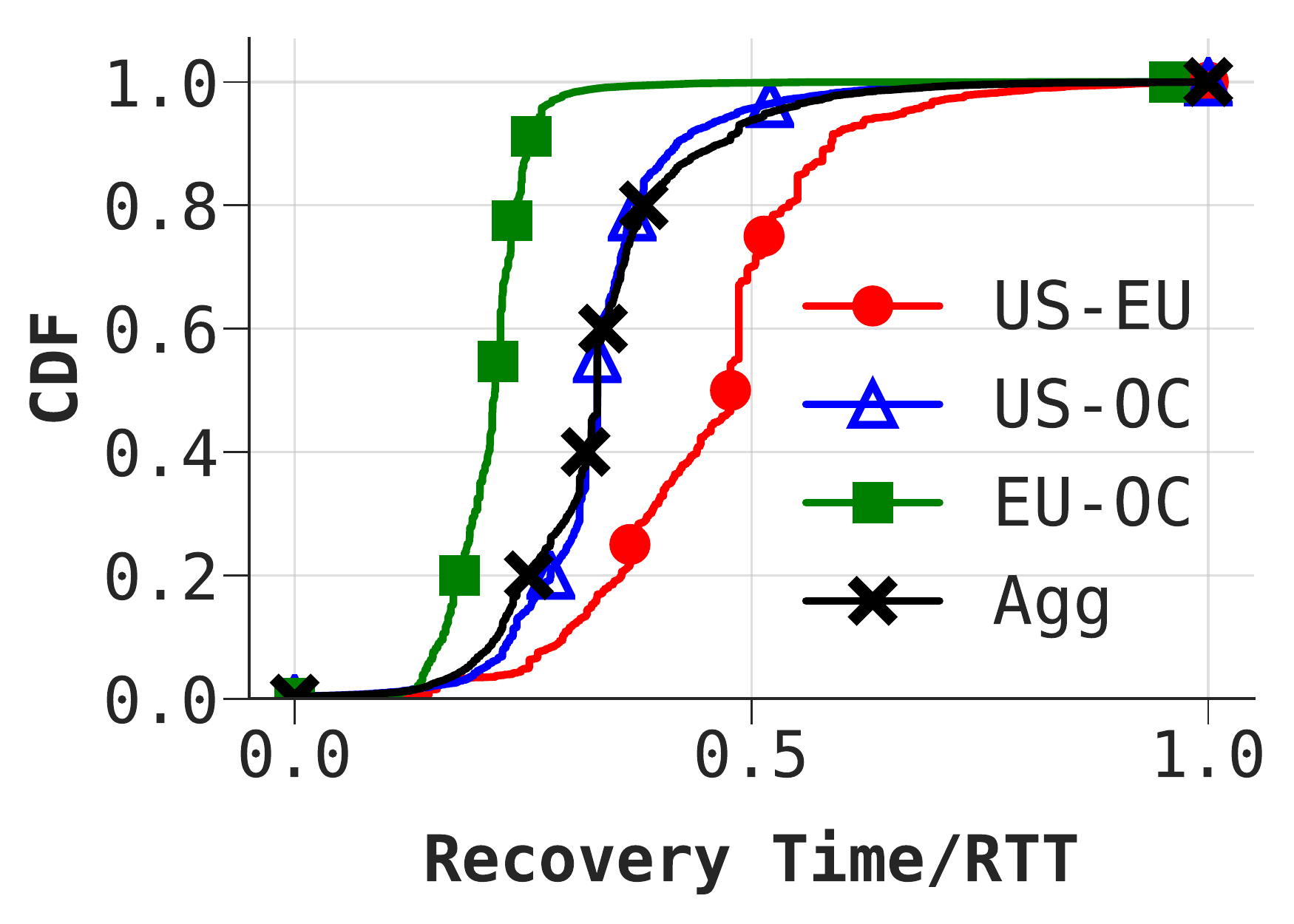}\label{fig:rtime_rtt}}
 \subfigure[]{\includegraphics[width=1.37in]{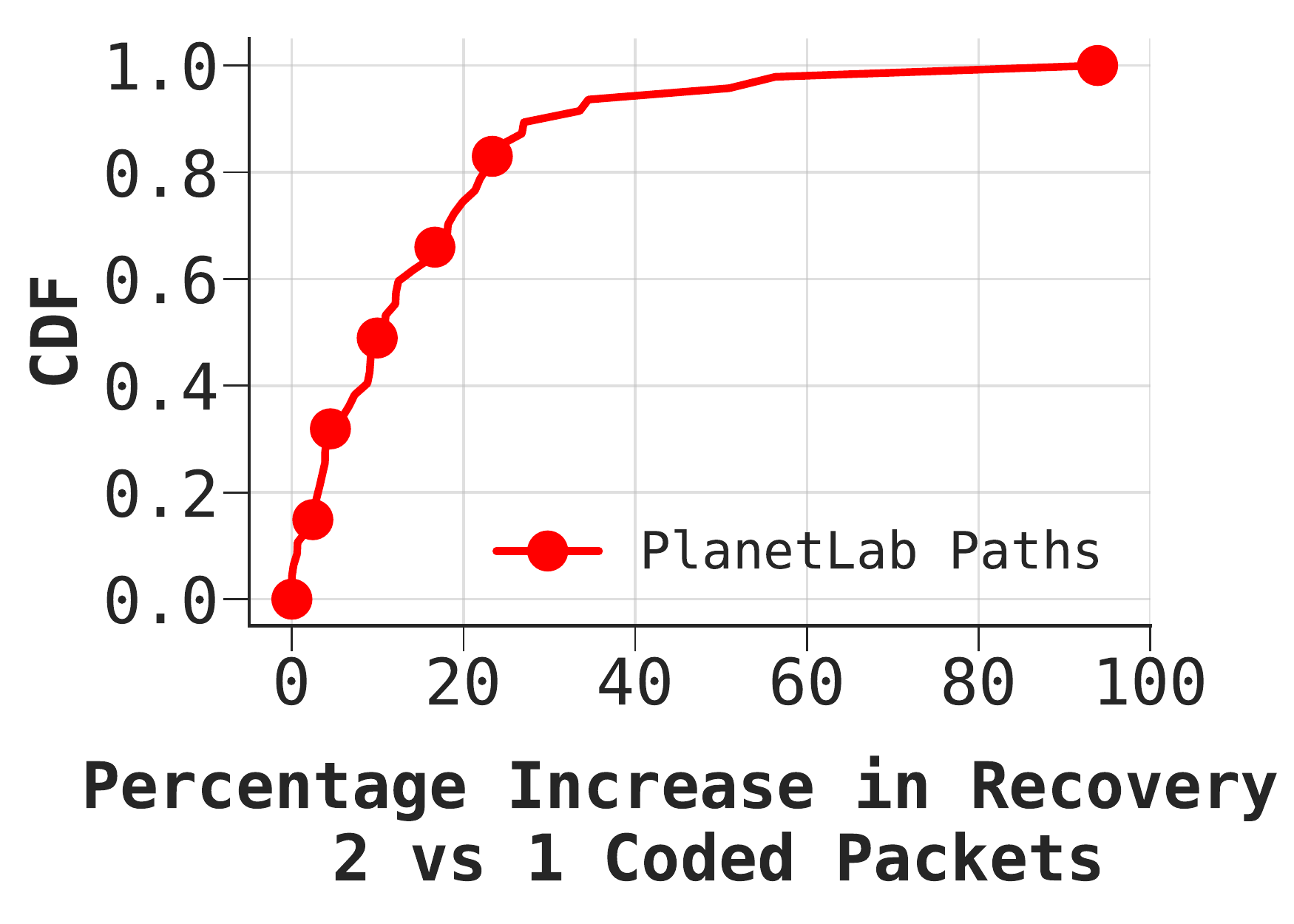}\label{fig:reco_diff}}

  \caption{\rewan{}'s performance on PlanetLab paths.(a) CCDF of successfully recovered packets. (b) Loss episode contribution to loss rate on paths with greater than 80\% recovery. (c) Percentage increase in \rewan{} recovery rate vs FEC with 20\%, 40\%,and 100\% packet overhead on direct path (note x-axis). (d) Packet recovery times as ratio of direct path RTT. (e) Percentage increase in recovery rates using 2 cross coded packets per batch versus 1. }
\label{fig:rewan-all-metrics}
 \end{figure*}

\subsection{\rewan{} Deployment and Evaluation}
\label{subsec:pl-eval}
We have evaluated \sys{} services under various controlled settings, verifying 
their ability to handle different uses which we described earlier. 
In this section, we evaluate our coding use case, \rewan{}, with its deployment and evaluation on the PlanetLab testbed. We select this service because it builds on the other services and adds the most delay in terms of packet recovery, representing the worst case scenario for
the effectiveness of \sys{} services.

\paragraph{Experimental Setup.} We ran \rewan{} as a service on five different DCs of Microsoft Azure~\cite{azure}, located in US (East, West), EU, Asia, and OC, for over a month. We use F1 type virtual machine, which is compute-optimized with 2.4 GHz single core and 2 GiB RAM. 
We evaluate 45 PlanetLab wide area paths spanning four different continents\footnote{Path details at \url{http://tinyurl.com/pl-paths}}.

We run a simple constant bitrate application on the PlanetLab nodes. 
To observe long-term time-averaged behavior without overloading the paths, we use ON/OFF periods with Poisson OFF times and constant ON times. In each ON interval, we send packets for 5 minutes with 10ms frequency; we set the mean OFF time to be 55 minutes.
DC1 relays the start of each ON interval to senders using a separate control channel, thereby ensuring that 
senders are (loosely) synchronized. We use $r=2/6$ and $s=1/5$ as our coding parameters.
Given the high churn rate of PlanetLab nodes, the total samples collected from each path varies.  Typically, we recover 500-800 samples per path, which translates to 3-5 weeks of measurement collection.
Our wide-area evaluation makes five key findings, summarized below:

\paragraph{Most losses happen on wide-area links and \rewan{} is able to recover them.} 
\rewan{} is able to recover 78\% of all packets that are lost on the PlanetLab paths. Loss rates on these paths are relatively high: up to 0.9\% loss, with 40\% of paths having a loss rate greater than 0.1\%.  Overall, we lose 0.02\% packets in our experiment and we consider any packet that takes longer than one RTT to recover as a lost packet. As we discuss later, most of the packets that \rewan{} is unable to recover are lost on the access paths. If we ignore those losses, \rewan{}'s packet recovery goes up significantly. 

Figure~\ref{fig:rewan_recovery_overview} elaborates on the above results -- it shows a CCDF of the fraction of successfully recovered packets (i.e., those lost packets that are recovered within one RTT) for all PlanetLab paths. Most paths experience high recovery (low unrecovered packet rate) -- overall, 82\% of paths successfully recover more than 80\% of lost packets.

\paragraph{\rewan{}'s coding is able to handle a wide range of loss patterns.} We next 
zoom into the loss patterns to understand what types of losses are being recovered by \rewan{}. Figure~\ref{fig:rewan_burst_overview} shows a CDF of loss episode patterns observed on PlanetLab paths that have greater than 80\% packet recovery (82\% of total paths). 
We look at the burst length of the loss episode and classify them as Random (single packet loss), Multi-Packet (2-14 packets), and Outage ($>$14 packets). We observe all three types of loss patterns on the chosen paths. While random and multi-packet bursts contribute more towards the loss rate, outages are not uncommon on these paths. Our data shows that 45\% of paths see outages that last from 1 to 3 seconds. Our recovery rates show that \sys{} service can handle multiple types of burst lengths, quickly.

\paragraph{Most access losses can be recovered using existing techniques.} 
While access losses (between source-DC1 and DC2-receiver) are not the main focus of \sys{}, we look at their loss characteristics to see whether well-known techniques can be used to recover such losses.  Our results show that around 98\% of such losses occur on source-DC1 paths and that a significant fraction, 90\%, of loss bursts are single packet losses and can be recovered using simple retransmissions (ARQ) or other simple redundancy based techniques (e.g.,~\cite{gentleaggression}) at the edges (i.e., between the end-points and the DCs).  In future, we plan to augment \sys{} to incorporate this observation.

\paragraph{\rewan{} vs. On-Path FEC schemes.}
To compare \rewan{} with traditional, on-path FEC packet recovery schemes, we perform a what-if analysis on the probes sent on the direct PlanetLab paths.
Our goal is to compare \rewan{} with sending different number of FEC packets on the direct path. 
We divide the probes into 5 packet bursts and consider the next burst as the FEC packets. 
We then compute recovery success rates for 20\% ($s=\frac{1}{5}$), 40\% ($s=\frac{2}{5}$), and 100\% ($s=\frac{5}{5}$) FEC overhead. We also assume that, for \rewan{}, access losses can be recovered using existing 
techniques. 

Figure~\ref{fig:rewan_fec_diff} shows the percentage increase in recovery rates for all the paths using \rewan{}, compared to different levels of FEC. 
We observe  that even at 100\% overhead (full duplication), 90\% of the paths had at least one loss episode
that could have been recovered using \rewan{} but not with on-path, 100\% FEC overhead. 
Further, 10\% of the paths observe more than 160\% improvement in recovery rates with \rewan{} compared
to full, on-path duplication. These are paths that experience long burst of losses or outages that cannot be recovered using FEC on the direct path.  For 20\% overhead scheme, 100\% increase in recovery rate is seen by 70\% of the paths. This result shows that there exist paths for which \rewan{}'s cross-stream coding is more effective in recovering from outages and bursty losses compared to traditional, on-path FEC based schemes.

\paragraph{\rewan{}'s loss recovery is usually fast.}  
We next look at packet recovery time using \rewan{}, which  Figure~\ref{fig:rtime_rtt} depicts for paths in different regions. We show our recovery times as a ratio of direct public Internet path RTT between the source and destination. We note that 95\% of packets are recovered within 0.5 $\times$ RTT. As expected, we observe faster recovery for paths with higher absolute latency on the direct public Internet path. For example, on low RTT paths between the US and EU (110-130 ms), we see higher recovery times as a proportion of RTT, but in terms of absolute latency, 90\% of packets are retrieved within 75 ms. We also observe that receiver-DC2 RTTs on these paths vary significantly. For example, the RTT between receivers in the EU and their nearest data center varies from 16-70 ms ($\mu$ = 28 ms).

However, as cloud providers continue to strive towards reducing their latency to end-users~\cite{mappinggoogle}, we expect \rewan{} recovery times to continue to improve over time.

Finally, we observe two systematic reasons contributing to the tail in the recovery time (Figure~\ref{fig:rtime_rtt}): delay in detecting and recovering a loss (e.g., due to delayed NACKs) and delay in arrival of coded packets at DC2. Overall, the percentage of recovered packets that fall outside of a reasonable time budget value is low and only accounts for roughly 1\% of the recovered packets.

\paragraph{Recovery time is improved due to straggler protection.}

Last, we show the benefit of using extra cross-stream coded packets to provide protection against stragglers during cooperative recovery.
Figure~\ref{fig:reco_diff} 
shows the performance gains using two cross-stream coded packets per batch, as opposed to one. 
We observe that, with adequate protection of two packets per batch, 60\% of paths see greater than 10\% improvement in recovery rates.  
We also observe that the recovery times decrease by at least 50 ms for 70\% of the recovered packets (not shown) -- in some instances, the difference is some stragglers that take several seconds.  
This further justifies our choice of default parameter values for PlanetLab paths.

\subsection{Case Study: Skype Performance and Cost}
\label{subsec:skype-perf}

We run \sys{} services under Skype's video conferencing scenario to measure their interaction with a popular, interactive application.
We focus on the performance of Skype in wide-area settings where outages occur (similar to ones described earlier in our wide-area evaluation). To do so, we leverage the cloud path to run the video conference in three experiments. First, we examine how the video quality degrades during an outage along a public Internet path used by Skype. We then duplicate \textit{all} Skype packets over a cloud path (\sys{}'s forwarding service) to show that such a path can indeed make up for lost packets during outages. Finally, we use \rewan{} to selectively transmit coded packets over the cloud path and perform recovery at the receiver.

\paragraph{Testbed and Measurement Procedure.} We use a similar testbed to that used by Zhang et al.~\cite{zhang2012profiling}, in which clients communicate using Skype's video conferencing service. We connect clients running Skype for Linux 4.3 in a LAN, and emulate wide area path characteristics such as latency, packet loss rate, and jitter.

We use Skype's screen sharing mode to transmit a pre-recorded video that closely represents the normal motions of human interaction during a video conference. We then compare the quality of each received video against the reference video by converting all videos to raw (uncompressed) format, and compute objective QoE scores on a frame-by-frame basis using VQMT~\cite{vqmt}. Although objective video quality metrics are not as reliable as subjective metrics given by users (such as Mean Opinion Score), they are sufficient to approximate the quality of the video on a frame-by-frame basis. We show the scores of each frame in a CDF to approximate the quality of each video in aggregate.

\paragraph{Use of the forwarding sevice enables higher QoE.} Figure~\ref{fig:rewan_skype_eval} shows the video quality results as we vary the network conditions and paths used. When a 30 second outage occurs along the Internet path, Skype's built-in FEC mechanism is insufficient to maintain an acceptable level of QoE. The video quality degrades with pixelation and frozen video, and the number of frames with poor PSNR scores significantly increases. Due to the high availability of the cloud path, when we use the forwarding service during the 30-second Internet path outage, virtually all packets reach the destination, preserving the video quality (similar to an Internet path with a 0\% loss rate). This shows that Skype is amenable to using \sys{} services running in tandem with it to correct losses on its direct public Internet path.

\paragraph{\rewan{} achieves similar QoE compared to the forwarding service.} When running Skype over \rewan{}, we disable in-stream coding on the cloud path ($s = 0$), since Skype uses its own FEC techniques on the Internet path to recover lost packets~\cite{chitchat}. To use cross-stream coding, we inject three \textasciitilde200 Kbps background UDP flows whose packets are coded with Skype packets at DC1 at a rate of $r = 1/4$, with $k = 4$. Figure~\ref{fig:rewan_skype_eval} shows that \rewan{} achieves a similar level of QoE compared to using the forwarding service. 

\paragraph{\rewan{} uses significantly less bandwidth than the forwarding service.} As Skype uses its own FEC, we only need to utilize cross-stream coding, reducing the amount of inter-DC bandwidth used. We also observed the inter-arrival time of packets during Skype calls, and tuned \sys{} accordingly by setting the NACK timeout value to 25 ms. This reduces the number of false positive NACKs that trigger unnecessary cooperative recovery. By taking advantage of this application-specific knowledge, \sys{} achieves similar QoE scores as the forwarding service but uses much less bandwidth: in our experiments, \sys{} sent just 13.4\% as many packets and 13.6\% as many bytes as did the forwarding service.

 \begin{figure}[!t]
  \subfigure[]{\includegraphics[scale=0.11]{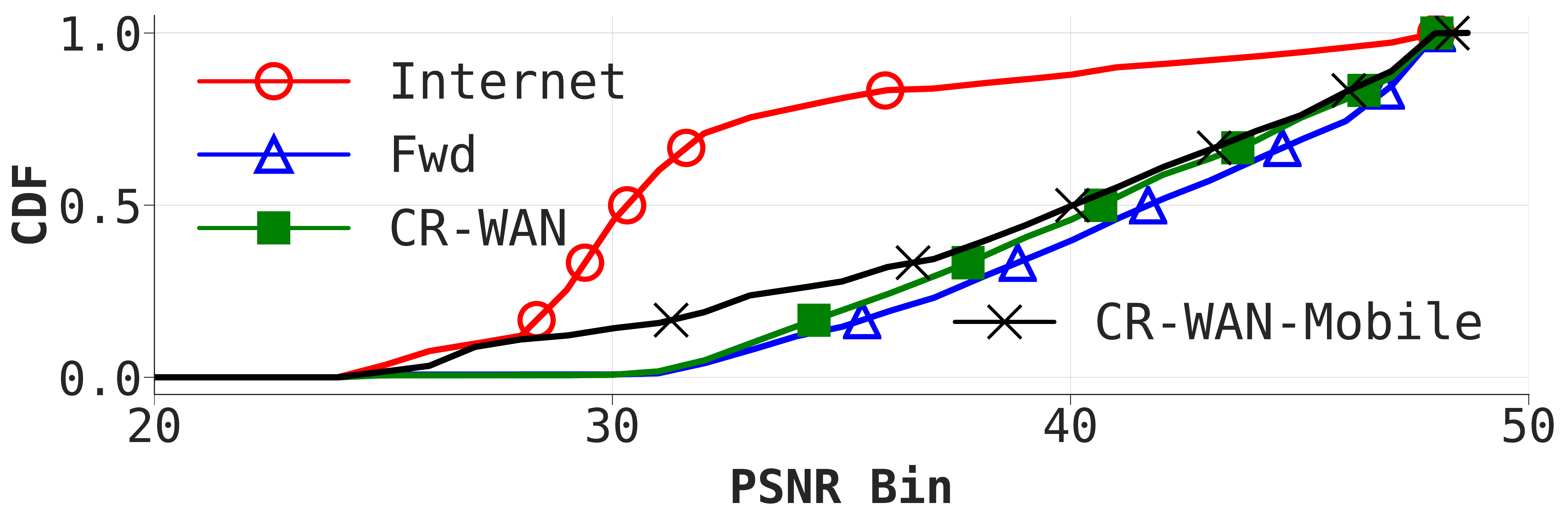}\label{fig:rewan_skype_eval}}
  \hspace{-0.60em}
  \subfigure[]{\includegraphics[width=1.05in]{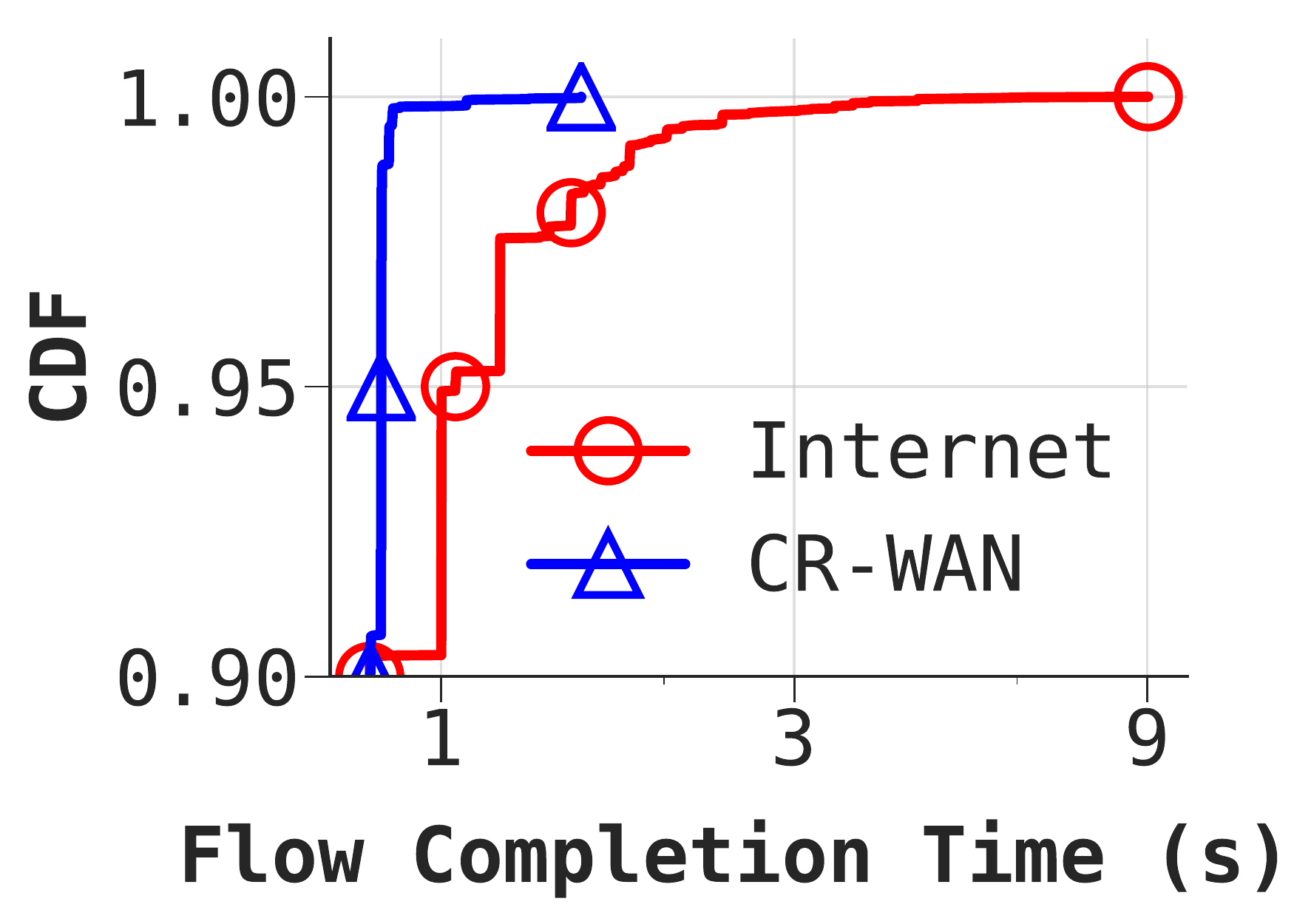}\label{fig:rewan_tcp}}
  \caption{a) PSNR scores of a set of video conferences,  b) Tail of TCP flow completion times (note y-axis scale).}
  \label{fig:rewan-s-and-t}
 \end{figure}

\subsection{Case Study: TCP Performance}
\label{subsec:rewan-tcp-eval}
We now evaluate the performance of TCP if it is used over \sys{}. Our goal is to understand how the additional reliability provided by \sys{} interacts with TCP's own reliability and congestion control mechanisms, and whether it can provide any additional benefits. 
We also evaluate whether we can use \sys{} services for only some (selective) TCP packets rather than all the packets. 
We focus on TCP short flows as they are latency sensitive and do not require high throughput.

\paragraph{Experimental Setup.} Our experimental setup is inspired by a similar experiment conducted by Google to evaluate different loss mitigation techniques for their web transfers~\cite{gentleaggression}. Using Emulab, we emulate the same topology and loss model as used in the Google study:  we consider a 200 ms RTT between end hosts and loss probabilities of 0.01 for losing the first packet in a burst and 0.5 for each subsequent loss. We pick \rewan{} for our analysis as it incurs the highest delay out of all the \sys{} services.  
We consider a single client-server scenario, in which a client sends a 12B request and receives a 50 KB response from the server. The RTT between server/client-DC paths is 30 ms with an RTT of 200 ms on the DC1-DC2 path. We make 10K requests each for TCP and TCP over \sys{}.

\paragraph{\sys{} reduces tail latency for lossy short flows.}
Figure~\ref{fig:rewan_tcp} shows TCP's flow completion times with and without \sys{}. 
We observe that shows that TCP suffers from long latency tail that goes up to 9 seconds, whereas \sys{} reduces the tail significantly. Our analysis shows that TCP is able to recover from most of the losses (using SACK), but there are some losses which are problematic for TCP, and hence cause the long tail. Such losses typically occur at the start of the connection, e.g., SYN-ACK(s), or at the very end. Such losses cause 
TCP to timeout, and successive losses mean that these timeout values could become huge, resulting in the long tail for TCP. 
\sys{} is able to reduce flow completion times by quickly recovering these losses. As soon as a packet is recovered by \sys{}, our TCP client sends an ACK to the server, effectively hiding the loss, and avoiding TCP timeouts.

\paragraph{Selective duplication can yield some benefits.}
When full duplication at source is infeasible -- due to limited access bandwidth or applications with high bitrates -- we can use \sys{} only for selected packets. To demonstrate the feasibility of such a strategy (and its potential benefits), we modify our TCP experiment and only duplicate SYN-ACK packets. We observe that selective duplication reduces tail by 33\% (83\% with full duplication). Other examples of such duplication can include I-frames for video streaming, important user actions for gaming or AR applications, and the last packet of a window for short TCP transfers~\cite{gentleaggression}

\subsection{Case Study: Mobile Networks}
\label{subsec:mobile-eval}
Some of the \sys{} services make assumptions that can be challenged in mobile networks, as mobile settings have different bandwidth, power, and latency characteristics. We pick \rewan{} for our analysis as it subsumes other services, in terms of its overhead. 
Our findings suggest that while it seems feasible to run \rewan{} on mobile hosts, it may be best to use selective duplication to avoid extra overheads.

\paragraph{Duplicating traffic can be feasible.} 
The bandwidth provided to cellular devices can vary greatly~\cite{sprout2013} -- our survey of major US carriers shows users can typically expect 2-5 Mbps uplink bandwidth. Therefore, we consider whether the most bandwidth intensive part of \rewan{} -- the duplication of traffic to the cloud path at the sender -- works within the link rates of mobile networks.

We modified our Skype testbed (\S\ref{subsec:skype-perf}) to tether the sending host to a mobile device connected to an LTE network, and observed that the overall bandwidth required by \rewan{} to duplicate a Skype video stream was 1.5 Mbps and well within the uplink bandwidth 
afforded by the LTE network ($\sim$5.0 Mbps). However, in general the recommended bandwidth for HD video calls in Skype is 1.5 Mbps~\cite{skype_bw}, so duplicating that traffic to over 3.0 Mbps could reach the capacity of uplinks in some networks. We also tested how \rewan{} affects other ongoing transfers on the device, and found that the transfer time for 5 MB files over WhatsApp is not affected by \rewan{} running simultaneously.

For data-intensive uses, \sys{} may need to utilize the forwarding service so that packets are not duplicated. Alternatively, mobile applications might selectively duplicate packets when using caching or coding and the Internet path performance is below a certain requirement. While current access capacity limits the use of \sys{} services, cellular bandwidth is expected to increase in future with 5G networks.

\paragraph{Duplicating traffic has negligible impact on power consumption.} We tested the effect of duplicating a traffic stream on the battery life of the device. We ran 20 minute trials of Skype video calls, with and without cloud path duplication. We observed that in both cases the battery drain was $\sim$20 mAh, highlighting that the extra overhead of \rewan{} has negligible impact on battery life. 

\paragraph{Recovery can be feasible despite latency issues.} Mobile networks also suffer from greater end-to-end latency and jitter~\cite{sprout2013}. We conducted a short study to quantify this effect by pinging three major cloud providers (Amazon, Microsoft, and Google) 1,000 times using different mobile networks: Verizon's LTE network (east coast) and T-Mobile's LTE network (both east and west coasts). 
The median ping times to each provider was typically in the range of 50-60 ms, but the 50\%-90\% RTTs to each cloud provider was in the range of approximately 50-100 ms.

These latencies could be problematic for mobile receivers, as the effect of greater latency is multiplied during recovery, especially during the coding service's cooperative recovery process. Despite this, our mobile Skype testbed was able to recover packets during an outage (fig~\ref{fig:rewan_skype_eval}) because the application is able to adapt to a greater end-to-end delay as long as it is consistent. In addition, due to increased jitter, correcting random packet losses may be difficult for interactive applications, but can likely be mitigated for other applications (such as web transfers) using in-stream coding. Finally, with cellular latencies expected to go down with 5G networks, 
the recovery delays will become smaller in future.

\section{Related Work}
\label{sec:related-work}

\sys{} connects to and benefits from a large body of prior work. We comment on key pieces from the literature that are most relevant to our study.

\noindent\textbf{Overlay Networks and Internet Architectures.} Our work is inspired by \emph{overlay} networks that improve availability by using detour points, e.g., RON~\cite{ron}, OverQoS~\cite{overqos}, one-hop source routing~\cite{detour}, Spines~\cite{spines}, etc. 
Our use of the cloud as an overlay creates unique opportunities and challenges. For example, we can do sub-RTT recovery, but to minimize cost, we have to send an additional small number of recovery packets. 
Recently, there have been proposals that make the case for using cloud as an overlay to improve QoS of interactive applications ~\cite{via2016,rewanhotnets,babay2017timely},  TCP-based applications~\cite{cronets2016,piedpiper-arxiv}, and provide Network-as-a-Service~\cite{movingbits-ifip2018}. Schemes like VIA~\cite{via2016} improve performance by routing \emph{all} of a certain user's traffic through the overlay path whereas ReWAN~\cite{rewanhotnets} provides a high level idea of using coded packets across the cloud paths for packet recovery. While we take into account cost of using cloud infrastructure and propose different use cases that use cloud storage and processing (e.g. coding across streams, caching), other schemes~\cite{cronets2016,piedpiper-arxiv,babay2017timely} route \emph{all} of the traffic through cloud overlays, similar to our forwarding service. 

\noindent\textbf{Caching.} Our caching service is similar in spirit to various Internet architecture proposals that do in-network caching e.g. NDN~\cite{Jacobson09}, i3~\cite{i3}, XIA~\cite{xiansdi}, etc. A DC that stores packet for later delivery can be thought of as a rendezvous point as in i3~\cite{i3}, or as a fallback host like XIA~\cite{xiansdi}.
Traditionally, CDNs are also used to store and deliver content.
RPT(~\cite{rpt-nsdi12}) is the most relevant as it uses caching for loss recovery -- it uses on-path content aware routers to compress and decompress packets on every hop. 
Our use of caching is unique as we use it \emph{only} for packet recovery: we use nearby off-path DCs to storage packet and receivers pull lost packets from the nearby cache.  Further, we only store a flow's packet for very short times (1-2RTTs) whereas typical caches (e.g., CDNs) store content for longer duration and serve multiple users. 

\noindent\textbf{Coding.}
Traditionally, network coding techniques have seen widest use in  the context of wireless networks~\cite{Katti06,networkcoding2006}. 
\sys{} applies cross-stream coding on wide area Internet paths and uses it to recover lost packets.
FEC based coding schemes have also been used in different contexts over the last several decades. The most relevant work to our scheme is Maelstrom~\cite{maelstrom}, which uses an FEC-based technique to reduce packet loss on lambda networks. Maelstrom's layered interleaving provides additional protection against bursty losses, but at the expense of higher decoding delay, which limits its use for highly interactive applications. Also, unlike Maelstrom,  the coded and data packets are sent on \emph{different} paths, with very different properties.   

\noindent\textbf{Reliable and Low Latency Wide Area Communication.} Finally, we share the goals of recent proposals that call for \emph{low latency and high reliability} for wide area communication~\cite{speedoflight, arrow,singla-cisp, coloc-imc17}. 
Some recent proposals focus on improving performance of TCP short flows~\cite{ietf-tcpm-rack, tcpm-tcp-loss-probe} using different techniques (e.g. per packet timestamps, early retranmissions). We, however, use the nearest DC to recover lost packets within a fraction of a path's RTT.

\section{Conclusion}
\label{sec:concl}

\sys{} seeks to connect two complementary interests: the \emph{pull} of existing (and burgeoning) applications and their demand for better user experience, and the \emph{push} of DC technology that makes cloud services more accessible to the edge than ever before. 
The key idea behind \sys{} is to use the cloud paths in a judicious manner, in order to provide reliability services to applications with different latency requirements. 
We view \sys{} as a promising step toward providing application and network architects with new insights into how to judiciously leverage the cloud.

\newpage

\section*{Acknowledgments}
We thank the anonymous reviewers for their feedback on this work, and Shawn Doughty for helping with PlanetLab access. This work was partially supported by NSF CNS under award numbers 1815046 and 1815016.

\setlength{\bibsep}{2pt plus 1pt}  
\small
\bibliography{ref}
\bibliographystyle{abbrvnat}
}{
}

\end{document}